\documentclass{aa}

\usepackage[T1]{fontenc}

\DeclareRobustCommand{\VAN}[3]{#2}
\let\VANthebibliography\thebibliography
\def\thebibliography{\DeclareRobustCommand{\VAN}[3]{##3}\VANthebibliography}

\usepackage{graphicx,txfonts}
\usepackage{subcaption}
\usepackage{placeins}    
\usepackage{siunitx}		
\usepackage{amsmath}	
\usepackage{amssymb}	
\usepackage{aas_macros}

\usepackage[shortcuts,acronyms]{glossaries-extra}	
\usepackage{orcidlink}
\usepackage{amsfonts}
\usepackage{float}

\begin{document}
\nolinenumbers

\title{Investigating the origin of the Milky Way streams}
\subtitle{A revised look at their orbital pole distribution in light of precession effects}

\author{
Elena Asencio\orcidlink{0000-0002-3951-8718}$^{1}$\thanks{E-mail: s6elena@uni-bonn.de (EA)}, 
Pavel Kroupa\orcidlink{0000-0002-7301-3377}$^{1,2}$
and Ingo Thies\orcidlink{0009-0001-5708-8088}$^{1}$}

\institute{
$^{1}$Helmholtz Institut für Strahlen und Kernphysik, Universität Bonn, Nussallee 1416, 53115 Bonn, Germany\\
$^{2}$Charles University, Faculty of Mathematics and Physics, Astronomical Institute, V Hole\v{s}ovi\v{c}kách 2, Praha, CZ-18000, Czech Republic\\
}

\date{Accepted XXX. Received YYY; in original form ZZZ}

\abstract 
{Stellar streams around the Milky Way (MW) can provide valuable insights into its history and substructure formation. Previous studies have suggested that several MW streams could have an origin related to that of the disc of satellite galaxies (DoS) and the young halo globular clusters of the MW, given that many of these structures present a similar orbital pole orientation. In this work we test the validity of this hypothesis by revising the orbital pole distribution of the MW streams with the latest stream dataset (\textsc{galstreams}). For a sample of 91 streams at Galactocentric distances of $d<100$~kpc we find that the pole distribution has no preferred orbital direction. However, as we subtract the streams closer to the Galactic centre, by imposing several lower distance cuts, we find that the larger the Galactocentric distance of the streams, the higher the fraction of stream poles pointing in a direction similar to the DoS. This trend could be explained if the stream pole distribution were originally anisotropic, but precession effects displaced the orbital poles of the streams closer to the Galactic centre. From the pole distribution and the estimated precession rates of the streams in the sample, we infer that the streams nearer the Galactic centre are indeed quite likely to be affected by precession. Finally, we corroborate with hydrodynamical simulations that, even in a scenario in which the MW substructures had a common origin, an overdensity in their orbital pole direction cannot be appreciated until the selected sample also includes material at $d \gtrsim 150$~kpc.}


\titlerunning{Are the Milky Way streams consistent with the VPOS?}
\authorrunning{E. Asencio, P. Kroupa \& I. Thies}

\maketitle

\glsresetall

\nolinenumbers
\section{Introduction}
\label{Introduction}

The stellar streams of the Milky Way (MW) are understood to be disrupted remnants of star clusters and dwarf galaxies that are orbiting our Galaxy \citep{Kohler_2021, Bonaca_2025}. Several properties of the stellar streams have been shown to be influenced by the properties of their parent object. For instance, the mass and the density profile of a progenitor dwarf galaxy can determine the stream structure and its velocity dispersion \citep{Carlberg_2018, Malhan_2021}, and the orbit of the stars in tidal streams is similar to that of their parent object \citep{Odenkirchen_2003, Erkal_2016} $-$ unless the parent object is significantly affected by dynamical friction \citep{Pagnini_2023}. Therefore, stellar streams with similar properties are expected to have a related origin. Several studies have attempted to explain the origin of the MW stellar streams based on this \citep{Pawlowski_2012, Bonaca_2021, Li_2022}.

\citet{Pawlowski_2012} studied the MW streams based on the direction of their orbital poles. This approach was motivated by previous studies that had already found that the direction of the orbital poles of the MW satellite galaxies agreed remarkably well \citep{Kroupa_2005, Metz_2007}. \citet{Pawlowski_2012} found that 7 out of 14 streams presented a similar pole orientation to that of the disc of satellites (DoS) and the young globular clusters of the MW. This suggested a common origin for this group of structures, which the authors named the vast polar structure (VPOS).

Different scenarios have been proposed to explain the origin of the MW substructures. For instance, in the Lambda cold dark matter ($\Lambda$CDM) model \citep{Efstathiou_1990, Ostriker_1995} $-$ also known as the standard model of cosmology $-$ substructures are preferentially accreted into galaxies through the dark matter filaments that assemble the cosmic web \citep{Bond_1996, Lovell_2011}. The filamentary accretion is expected to cause some degree of anisotropy in the distribution of substructures orbiting the galaxies, but $\Lambda$CDM simulations have shown that this mechanism of accretion is insufficient to explain the highly correlated position and kinematics of the satellite discs observed in the MW, Andromeda (M31), and Centaurus A \citep{Chiboucas_2013, Ibata_2014, Pawlowski_2014, Pawlowski_2018, Pawlowski_2020, Pawlowski_2021, Muller_2021}. Other mechanisms, such as group infall, are only expected to bring groups of 3-5 dwarf galaxies into galaxies similar to the MW \citep{Julio_2024}, which would be insufficient to explain the correlation of the (more numerous) satellites in the aforementioned systems. Such groups of dwarfs are also expected to be more spatially extended than the satellite discs \citep{Metz_2009b}.

Given the difficulties of the previous scenarios in explaining the properties of the DoS, \citet{Pawlowski_2012} proposed an alternative model. In their scenario, the MW and M31 had an encounter in the past that led to the creation of tidal tails in these galaxies. Structures that originate from the same tidal tail are expected to be phase-space correlated \citep{Pawlowski_2011, Kroupa_2012, Pawlowski_2018, Haslbauer_2019, Bilek_2021}. Therefore, this scenario could explain the observed distribution of the dwarf galaxies in the DoS and the similarities in the orbital poles of the VPOS members.

An important implication of this scenario is that the dwarf galaxies of the MW would then be of tidal origin, and therefore dark-matter-free \citep{Barnes_1992, Wetzstein_2007}. In $\Lambda$CDM, dark matter is required to explain the high velocities of the objects orbiting the outskirts of galaxies and galaxy clusters, as these cannot be understood with Newtonian dynamics alone \citep{Oort_1940, Rubin_1970}. Additionally, the absence of a dark matter halo would make the galaxies more fragile and susceptible to be destroyed. According to \citet{Bournaud_2006}, only about 25\% of the dwarf galaxies formed in a tidal tail can become long-lived objects in $\Lambda$CDM. These would also be the most massive and distant dwarfs in the tail. Due to the large number and the spatial distribution of the satellite galaxies of the MW and M31 $-$ as well as Centaurus A $-$ it is unlikely that most of them have a tidal origin in this paradigm.

Beyond the $\Lambda$CDM model, there are other theories capable of explaining the dynamics of galaxies without resorting to dark matter. In these theories, the high velocity dispersion of galaxies is attributed to a non-Newtonian behaviour of gravity on galactic scales. The leading paradigm in this regard is currently Milgromian dynamics \citep[MOND]{Milgrom_1983}. In this model, gravity experiences a boost in the regime of low accelerations $-$ that is, when the gravitational acceleration goes below $a_0 = 1.2 \times 10^{-10}~\textrm{m/s}^2$ \citep{Begeman_1991, Gentile_2011}. Since in MOND the enhancement of Newtonian gravity is associated with an acceleration scale instead of dark matter, the high velocity dispersions of the satellite galaxies are still consistent with the model expectations, even if they are of tidal origin \citep{Brada_2000, Angus_2008, McGaugh_2010}. Moreover, in MOND, the absence of dark matter haloes around the MW and M31 makes it more likely for them to have a fly-by encounter without merging, as there is no additional dynamical friction coming from their haloes \citep{Tiret_2008, Zhao_2013, Kroupa_2015, Bilek_2018, Banik_2022}. For these reasons, MOND can be considered a more natural framework in which to understand the origin of the MW substructures.

However, a recent study by \citet{Riley_2020} found that, when using an updated streams catalogue with a significantly larger number of members (64) and the Gaia DR2 measurements for globular clusters \citep{Gaia_Collab_2018, Vasiliev_2019}, there is no preferential alignment of the orbital poles of the streams or the globular clusters in the VPOS direction. These results seem to challenge the hypothesis of a correlated origin for most MW substructures, including the MW-M31 fly-by scenario. \citet{Riley_2020} also commented on additional effects, such as precession, which could be affecting the direction of the orbital poles of the MW streams. But this was not taken into account in their analysis.

In our work we revisit the analyses performed by \citet{Pawlowski_2012} and \citet{Riley_2020} using the newest dataset on stellar streams \citep[\textsc{galstreams},][]{Mateu_2023}\footnote{\url{https://github.com/cmateu/galstreams/}} (Section~\ref{Obs_analysis}), which includes a total of 95 stream tracks with their corresponding orbital poles (see Section~\ref{Data} for a description of the dataset). We also compare the observed orbital pole distribution with the distribution resulting from a simulated MW-M31 interaction. For this purpose, we use the data from the MW-M31 simulation presented in \citet{Banik_2022} (Section~\ref{Sim_analysis}). We then discuss how precession effects could explain the currently observed orbital pole distribution, and the evolution of the pole distribution in the \citet{Banik_2022} simulation (Section~\ref{discussion}). In this work we do not repeat the analysis for the globular clusters, but the conclusions that we draw from our analyses can also be applied to the globular cluster pole distribution. In Section~\ref{conclusions} we summarise our main results.

\section{The stream poles}
\label{Data}
In order to obtain the orbital poles of the stellar streams that have been observed in the MW up to this day, we use the recently updated \textsc{galstreams} catalogue \citep{Mateu_2023}. This catalogue contains information on the angular position, distance, proper motion, and radial velocity track data of 95 stellar streams of the MW. The catalogue also includes the tracks of the streams, which \citet{Mateu_2023} reconstructed by fitting the empirical data available for these streams (e.g. reported knots, reference points, and stream members). The pole of a stream is expressed as a co-ordinate of its Galactic longitude and latitude $(l, b)$, and it is obtained as the normalised cross product of two spatial points in the stream \citep{Pawlowski_2012, Riley_2020}. This quantity is also provided in the catalogue both in heliocentric and Galactocentric co-ordinates. In our analysis, we use the Galactocentric orbital poles of the catalogue, for which the spatial points had previously been transformed to Galactocentric co-ordinates using the \textsc{python} library \textsc{astropy}. In its current version, \textsc{astropy} assumes a distance from the Sun to the Galactic centre of $d_{\odot-\textrm{GC}}=$~8.122~kpc \citep{Gravity_Collab_2018} and a configuration in which the x axis points towards the Galactic center from the projected position of the Sun in the Galactic midplane, the y axis points towards the Galactic longitude ($l = 90^{\circ}$), and the z axis points towards the north Galactic pole ($b = 90^{\circ}$).

The orbital sense of the poles can have two directions (clockwise or anti-clockwise). This direction is given by the radial and proper motions of the stream, which are still not available for most members of the catalogue. As in \citet{Pawlowski_2012} and \citet{Riley_2020}, we consider all the streams to have the same orbital sense, so we only need to account for half of the parameter space in the $l$ direction. For an easier comparison with the results of the aforementioned studies, we also choose the range $120^{\circ} < l < 300^{\circ}$.

Given that the streams were not formed in a perfectly spherical potential, their inferred orbital poles are generally not the same for every pair of spatial points along the stream. The asphericity of the MW potential can further exacerbate the differences in the angular positions of the stars forming the streams through precession and nutation \citep{Erkal_2016, Mateu_2017}. Besides this, the streams are also susceptible to experience other gravitational disturbances throughout their trajectory, for example the effect of massive satellite galaxies such as the Large Magellanic Cloud (LMC) \citep{Erkal_2019, Shipp_2019}, or the interactions with the Galactic bar \citep{Thomas_2023}. Because of this, the inferred direction of the stream pole is generally not the same for all the spatial points along the track. The \textsc{galstreams} catalogue provides, for each stream, the orbital poles obtained for all pairs of consecutive points in the stream track. These are referred to as `the pole tracks' \citep[see figure A2 in][]{Mateu_2023}.

\begin{figure*}
	\centering
	\includegraphics[width = 0.75\textwidth] {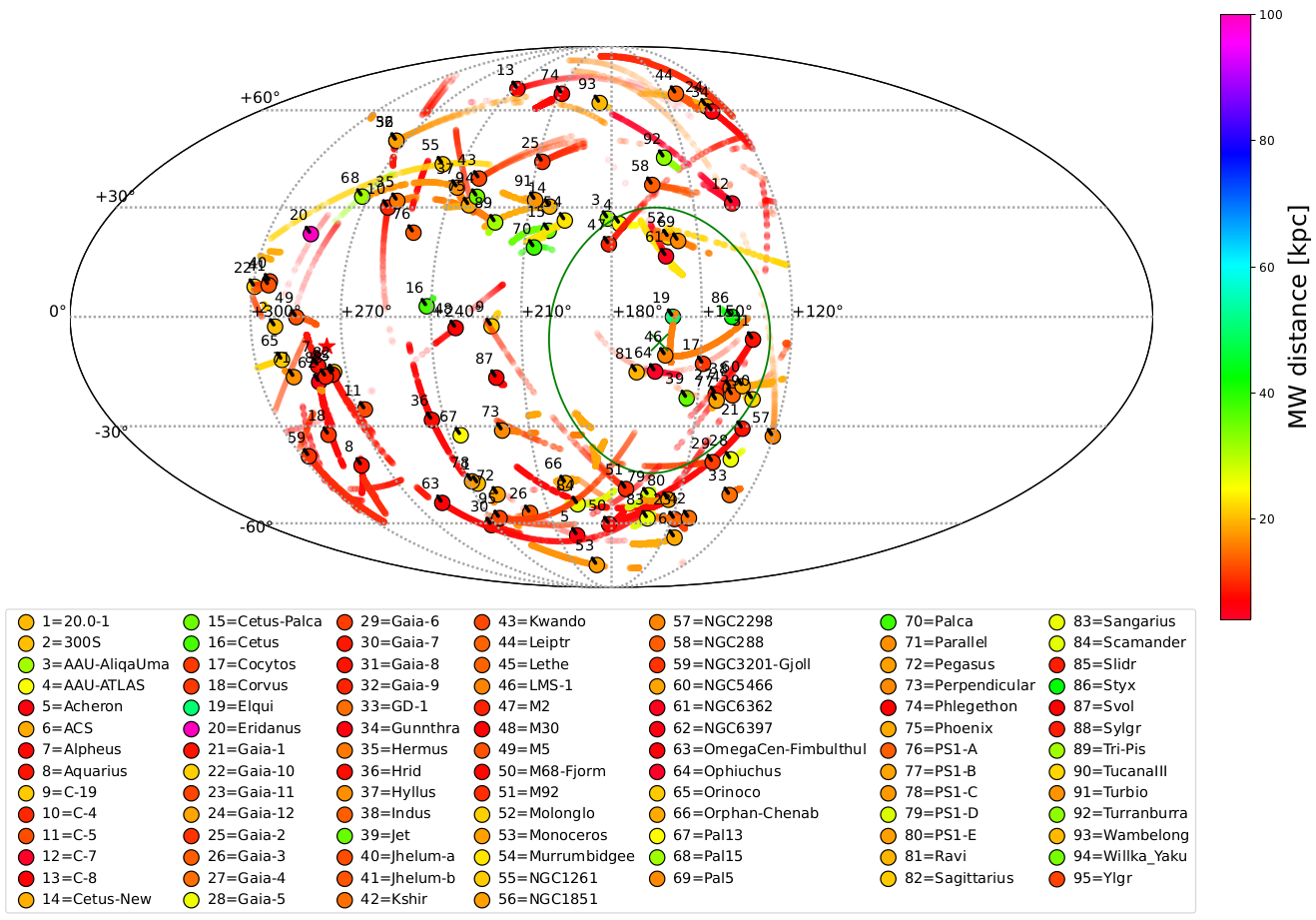}
	\caption{Distribution of the orbital poles of the MW streams included in the \textsc{galstreams} catalogue (represented with a Mollweide projection). Since we only consider one orbital sense, just one half ($120^{\circ} < l < 300^{\circ}$) of the $l$ range is shown. The nominal orbital poles for each stream are represented by the large coloured points. Their $1\sigma$ uncertainty is shown by the point clouds of the same colour. The green cross marks the position of the observed VPOS direction ($l$, $b$) = ($164.0^{\circ}$, $-6.9^{\circ}$) \citep{Pawlowski_2015} and the green circle surrounding it represents the threshold of the VPOS membership area considered in our statistical analysis (see Section~\ref{Obs_analysis}). The red star marks the position of the Sagittarius dwarf galaxy. The colour gradient of the streams represents their Galactocentric distance.}
	\label{poles_galstreams}
\end{figure*}

To find the most representative orbital pole in the track of each stream, we constructed a grid over the full parameter space of $2^{16}$ equal area bins using the \textsc{HEALPix}\footnote{\url{http://healpix.sourceforge.net}} \textsc{python} package \textsc{healpy} \citep{Gorski_2005, Zonca_2019}. We then counted the number of poles in the pole track that belong to each interval and we selected the bin with the highest incidence. Among the poles in that bin, we chose as the nominal pole the one that is closer to the mean of the $l$ and $b$ values in the bin. The rest of the poles in the pole track can be considered to be the uncertainty of the pole value. To give an estimate of the most likely values of the pole, we obtained the contour levels for the binned poles in the poletrack and we selected the pole values that are within the $1\sigma$ confidence region. The nominal orbital poles and their $1\sigma$ uncertainties obtained with this method for the \textsc{galstreams} catalogue are shown (under a Mollweide projection) in Figure~\ref{poles_galstreams}.

Each pole in a pole track also has a Galactocentric distance associated with it. For our study, we chose the distances corresponding to the nominal orbital poles as the nominal distances of the streams. As is shown in fig. 2 of \citet{Mateu_2023}, for most streams this quantity does not vary too much throughout the track. Therefore, we do not expect that the distance uncertainty will significantly affect our results. The streams in the catalogue have a distance range from 4~kpc to 100~kpc, but most of them (about $70\%$) are at a distance $\leq 20$~kpc (see Appendix \ref{App_distance}).

Some of the individual streams listed in the catalogue can be considered to be part of a single larger stream; for example, AAU-AliqaUma and AAU-ATLAS \citep{Li_2021}, Cetus-New, Cetus-Palca, and Cetus \citep{Yuan_2022, Thomas_2022}, and Jhelum-a and Jhelum-b \citep{Ibata_2021, Bonaca_2019}. As was expected, most of the orbital poles of the presumed substructures of the same stream have a similar pole direction. Given that in the following analysis we are assessing the degree of clustering of the poles, we chose only one stream for each of these groups (AAU-AliqaUma, Cetus, and Jhelum-a). In doing so, we expect to reduce the chances of overestimating the number of stream poles within a particular region of the parameter space. This leaves us with a total of 91 streams that we can use in our analysis.

Even though the current \textsc{galstreams} catalogue accounts for more recent data than the catalogues used in previous stream pole studies \citep{Pawlowski_2012, Riley_2020}, the stream pole positions for most of the streams have remained fairly similar (see Appendix~\ref{App_prev_studies}). Therefore, any potential differences in our results with respect to the ones of the previous studies are more likely to be caused by the increased number of streams in the catalogue.

\section{Statistical analysis of the observed pole distribution}
\label{Obs_analysis}
Similarly to \citet{Pawlowski_2012} and \citet{Riley_2020}, we investigated whether the stream poles have a preference for clustering near the VPOS normal ($l$, $b$) = ($164.0^{\circ}$, $-6.9^{\circ}$). The VPOS hypothesis proposes that most of the MW substructures (e.g. satellite galaxies, globular clusters, stellar streams) have a related origin, which should manifest as a preferred orbital pole direction in their pole distribution.

Presently, the normal of the VPOS structure is determined by fitting the orbits of the MW satellite galaxies, given that these substructures have already been confirmed to be part of a highly correlated orbital system by previous studies \citep{Pawlowski_2015, Pawlowski_2020}. For this study, we chose the VPOS normal inferred from the orbital poles of 38 satellite galaxies \citep[VPOS+new,][]{Pawlowski_2015}. The VPOS co-ordinates used in \citet{Pawlowski_2012} and in \citet{Riley_2020} were inferred from less recent samples, so their assumed VPOS was slightly different from ours. In Appendix~\ref{App_prev_studies}, we show that this does not have a significant impact on the results of the analysis.

As in \citet{Riley_2020}, we estimated that the substructures classified as VPOS members cover an area, around the VPOS normal, comprising $20\%$ of the considered hemisphere. This corresponds to a cone with an opening angle of $36.87^{\circ}$ (green circle in Fig.~\ref{poles_galstreams}). Then, we determined the number of stream poles that are enclosed within this area. Accounting for the uncertainty in the orbital pole direction, we only considered as VPOS members those streams whose $1\sigma$ uncertainty pole track has at least $50\%$ of its poles within a $36.87^{\circ}$ distance of the VPOS normal. To assess if the probability of observing stream poles within the selected area is affected by their distance to the MW, we classified the streams into VPOS members and non-VPOS members at the lower Galactocentric distance ($d$) cuts of $d>0$~kpc, $d>5$~kpc, $d>10$~kpc, and $d>15$~kpc. The results of this classification are shown in Table~\ref{tab_obs_statistics}. 

A preferential clustering of the stream poles around the VPOS direction will cause their distribution to deviate from a uniform one. In the next step, we quantified the probability of obtaining the observed fraction of VPOS members if the stream poles followed a uniform distribution $-$ so that lower probabilities would support distributions that favour the VPOS direction and vice versa. For this, we used the method described in \citet{Pawlowski_2012}, which can be summarised as follows: a Bernoulli experiment can give the probability of finding $k$ stream poles out of $n$ within a certain area, according to the assumed probability distribution; therefore, in order to find the probability that at least $k$ streams out of $n$ are within this area, one has to sum the probabilities of finding $k$, $k + 1$, ..., $n$ streams within this area. This gives us the total probability:
\begin{eqnarray}
\label{P_bern}
	P~=~ \sum_{i=0}^{n-k} \binom{n}{k+i} p(\theta)^{k+i} (1-p(\theta))^{n-k-i} \, ,
\end{eqnarray}
where $\theta$ is the aperture angle of the cone enclosing the selected area and $p(\theta)$ is the probability that an orbital pole is found within the selected area, given the assumed probability distribution. In this case, we have selected an aperture angle that encloses $20\%$ of the parameter space and a uniform distribution, which means that $p(\theta=36.87^{\circ}) = 0.2$. The resulting total probability, $P$, obtained for the previous VPOS classification and distance cuts, is included in Table~\ref{tab_obs_statistics}.

\begin{table}
	\caption{Number of VPOS members (in VPOS) and non-VPOS members (out VPOS).}
	\centering
	\begin{tabular}{c|c|c|c}
		Distance cut & in VPOS     & out VPOS            & $P$         \\ \hline
		\multicolumn{1}{c|}{$d>0$~kpc}     & 15 & 76 & $0.83 \pm 0.09$  \\ \hline
		\multicolumn{1}{c|}{$d>5$~kpc}     & 14 & 75 & $0.88 \pm 0.10$  \\ \hline
		\multicolumn{1}{c|}{$d>10$~kpc}    & 13 & 55 & $0.62 \pm 0.09$ \\ \hline
		\multicolumn{1}{c|}{$d>15$~kpc}    & 10 & 37 & $0.47 \pm 0.10$ \\ \hline
	\end{tabular}
	\tablefoot{The table also shows the total Bernoulli probability, $P$ (given by Eq.\ref{P_bern}), of observing such a fraction of VPOS members in a homogeneous distribution. The error of $P$ corresponds to the statistical uncertainty of the Bernoulli experiments ($\sqrt{P_i(1-P_i)/n}$ for each individual experiment in the sum).}
	\label{tab_obs_statistics}
\end{table}

When no distance cuts are imposed on the stream dataset, we find $P=0.83\pm 0.09$, which shows that the pole distribution is highly consistent with the expectations of a uniform distribution. We thus report a good agreement between our results and the results of \citet{Riley_2020}, who performed a similar analysis with a smaller dataset.

However, we also found a decrease in the $P$ values for the stream pole distribution at distances $d>10$~kpc and $d>15$~kpc. The decreasing trend of $P$ with increasing Galactocentric distance can be interpreted as the effect of precession causing the stream poles at low Galactocentric distances to tilt away from their originally anisotropic distribution. We discuss this possibility in further detail in Section~\ref{discussion})

\section{Hydrodynamical simulations of a MW-M31 interaction}
\label{Sim_analysis}

If the original stream pole distribution were indeed anisotropic but got dispersed over several gigayears, simulated MW substructures with an originally anisotropic pole distribution should have evolved similarly (at the Galactocentric distances of the observed sample). To test this, we used the results of a MONDian MW-M31 fly-by simulation \citep{Banik_2022} evolved up to the present day. In the simulation, the interaction of the two galaxies results in the formation of tidal debris, which, as was expected, presents a preferred orbital pole direction shortly after the interaction (see Appendix~\ref{App_t06}). In the present day, the preference for a similar pole direction is still maintained at large Galactocentric distances \citep[see][their fig.~7]{Banik_2022}.

Therefore, this data allows us to compare the observed stream pole distribution with the present-day distribution expected in the MW-M31 fly-by scenario $-$ that is, the scenario proposed by \citet{Pawlowski_2012} as the origin of the VPOS. Additionally, the simulation data can also help us to infer the Galactocentric distance at which the clustering of the orbital poles should become appreciable in the present day.

\subsection{The simulated model}
Among the MONDian simulations of a MW-M31 interaction performed up to the present day \citep{Zhao_2013, Bilek_2018, Banik_2022}, we chose the one by \citet{Banik_2022} for our analysis, as it provides information on the orbital poles of the substructures generated in the interaction. In order to perform their simulation, \citet{Banik_2022} used the $N$-body code \textsc{Phantom of Ramses} \citep[\textsc{PoR}][]{Lueghausen_2015, Nagesh_2021}. \textsc{PoR} is a MONDian version of the (Newtonian) $N$-body code \textsc{RAMSES} \citep{Teyssier_2002}. The MONDian patch allows the user to compute the gravitational potential of galaxies based on Milgrom's modification to Newtonian gravity \citep{Milgrom_1983}. In particular, \textsc{PoR} uses a version of this theory known as the `quasi-linear formulation of MOND' \citep[QUMOND,][]{Milgrom_2010}.

In their simulations, \citet{Banik_2022} set up two galaxies of $9.15 \times 10^{10}~M_{\odot}$ (MW) and $2.135 \times 10^{11}~M_{\odot}$ (M31) $-$ each consisting of $5 \times 10^5$ particles and a gas fraction of 50 percent. Their aim was to match the currently observed disc orientation and separation of the two galaxies, as well as the current orbital pole position of the satellite planes. For this, they set as constraints the present-day distance, direction, and radial velocities of the MW and M31, and varied their present-day tangential velocity to set up different models. For each of these, they explored the pole values of the tidal debris (particles and gas cells) resulting from the interaction. They chose as their best-fit model the configuration whose orbital pole overdensities best matched the observed satellite plane poles of the MW \citep[$l, b = (176.4^{\circ}, -15.0^{\circ})$, ][]{Pawlowski_2013} and M31 \citep[$l, b = (206.2^{\circ}, 7.8^{\circ})$, ][]{Pawlowski_2013b}. The orbital configuration corresponding to the best-fit model is consistent with the measured proper motions of M31, even if these were not set as a constraint in the simulation. In this model, the MW-M31 encounter took place $7.2$~Gyr ago.

\begin{figure*}
	\centering
 	\begin{subfigure}[b]{0.475\textwidth}
       \centering
       \includegraphics[width=\textwidth]{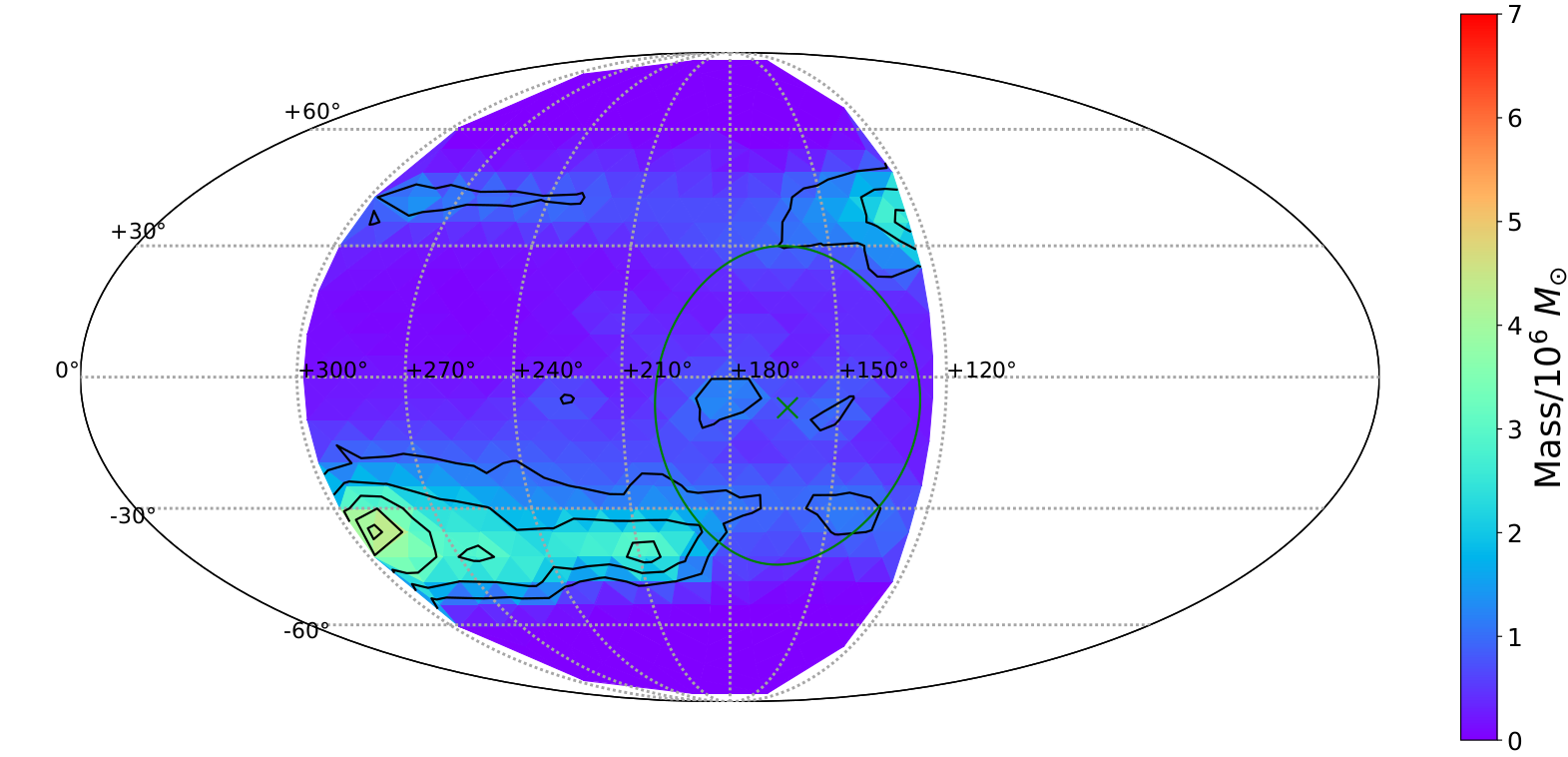}
       \caption[]{{\small Simulated tidal debris at $d<100$~kpc}}    
        \label{fig:sim_poles_1direction_d100}
     \end{subfigure}
     \hfill
     \begin{subfigure}[b]{0.475\textwidth}  
     	\centering 
        \includegraphics[width=\textwidth]{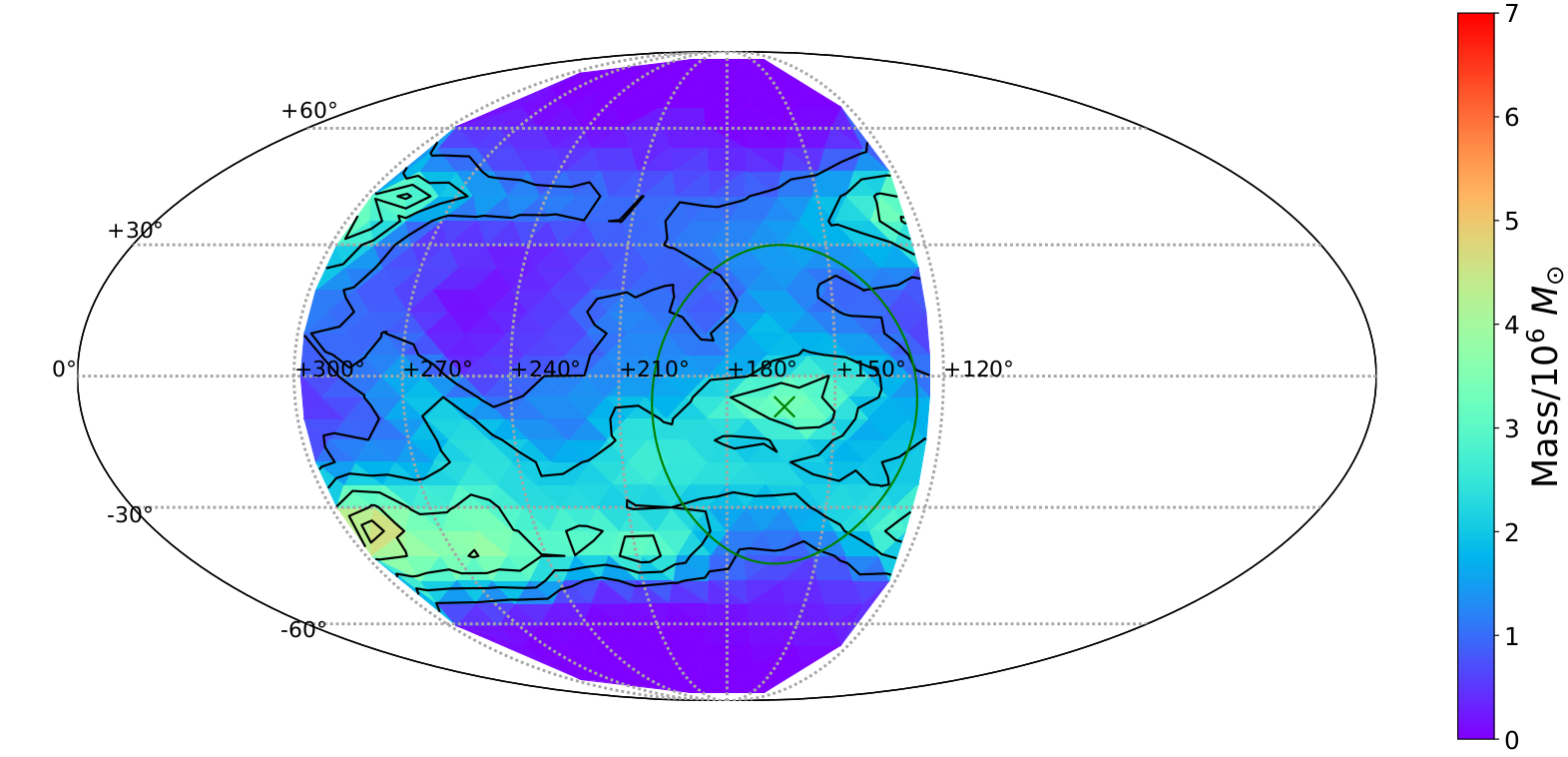}
        \caption[]{{\small Simulated tidal debris poles at $d<150$~kpc}}    
        \label{fig:sim_poles_1direction_d150}
    \end{subfigure}
     \vskip\baselineskip
     \begin{subfigure}[b]{0.475\textwidth}   
     	\centering 
        \includegraphics[width=\textwidth]{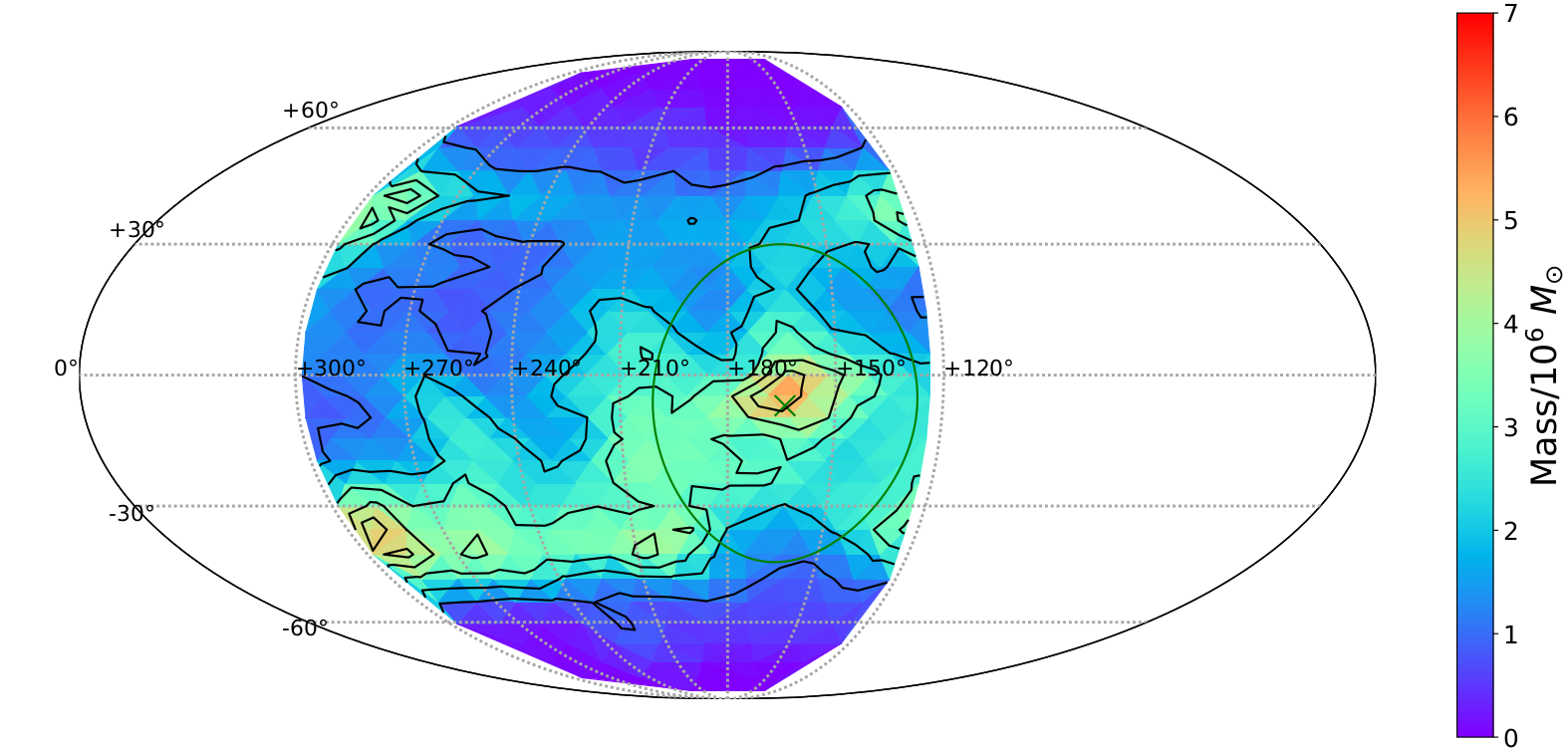}
        \caption[]{{\small Simulated tidal debris poles at $d<200$~kpc}}    
        \label{fig:sim_poles_1direction_d200}
        \end{subfigure}
        \hfill
	\begin{subfigure}[b]{0.475\textwidth}
        \centering 
        \includegraphics[width=\textwidth]{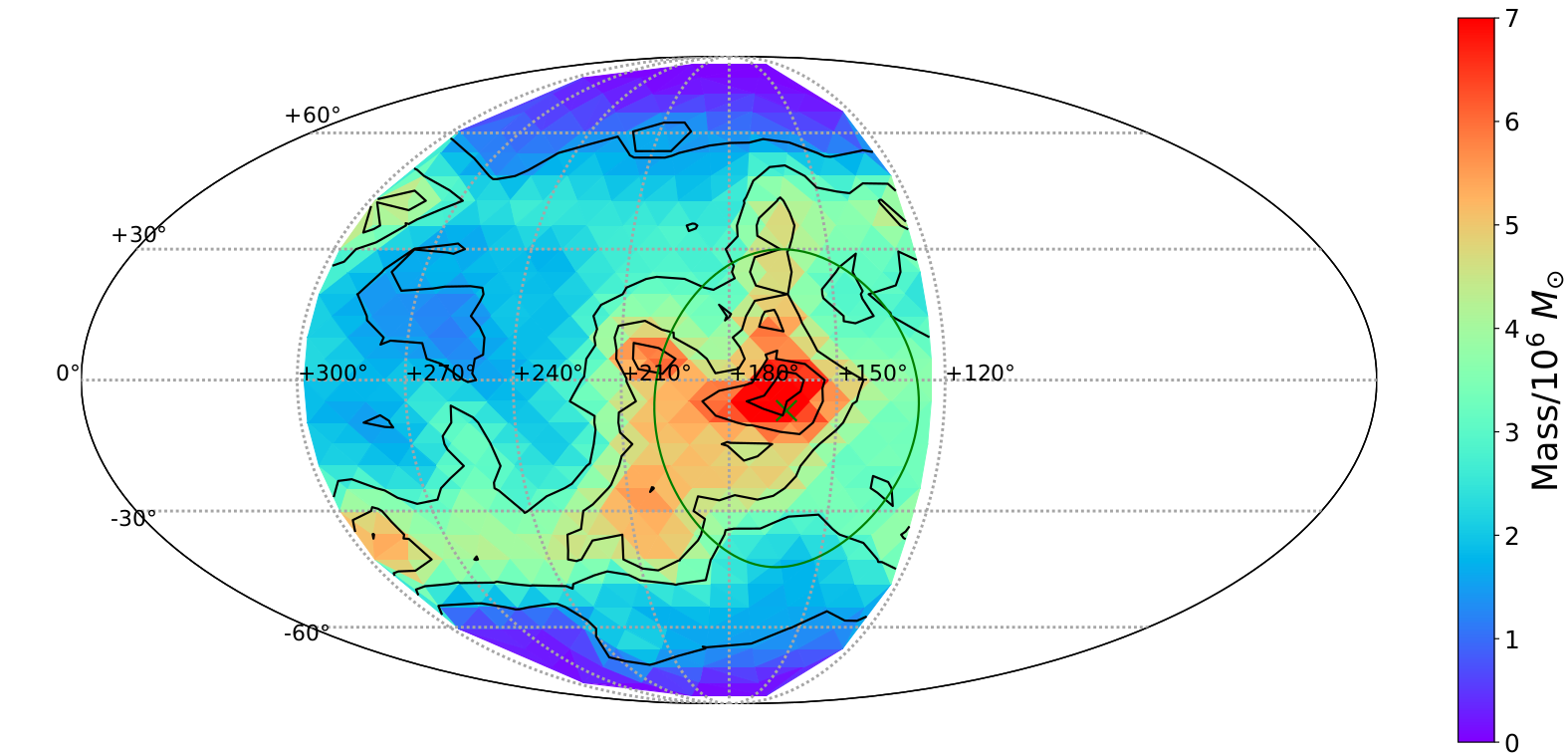}
        \caption[]{{\small Simulated tidal debris poles at $d<350$~kpc}}
        \label{fig:sim_poles_1direction_d350}
    \end{subfigure}
    \caption[]{Tidal debris pole distribution at different Galactocentric distance ($d$) cuts from the present-day \citet{Banik_2022} simulation data (represented with a Mollweide projection). All figures have a lower distance cut of $z > 50$~kpc to remove the contamination from the disc material. For a better comparison with the observed sample, we have also assumed that the simulated substructures are all orbiting with the same orbital sense (see Section~\ref{Data}), even though $-$ unlike the observed data $-$ the simulation does provide information on the orbital sense of its particles and gas cells. The green cross marks the position of the observed VPOS direction ($l$, $b$) = ($164.0^{\circ}$, $-6.9^{\circ}$) \citep{Pawlowski_2015} and the green circle surrounding it represents the threshold of the VPOS membership area considered in our statistical analysis (see Section~\ref{Obs_analysis}).}
    \label{fig:sim_poles_1direction}
\end{figure*}

In their simulation, \citet{Banik_2022} noted that, in order to avoid the contamination from MW disc particles and gas in the considered sample, the material that is at a distance of $z < 50$~kpc should be subtracted\footnote{Note that \citet{Banik_2022} use only the perpendicular component of the distance ($z$), with respect to the Galactic disc, to determine the distance cut at which the disc contamination should be mostly gone. For consistency, we also use just the $z$ component for distance cuts related to the removal of the disc contamination. For everything else, we continue to use the radial distance ($d$).}. In Fig.~\ref{fig:sim_poles_1direction} we show the simulated orbital pole distribution for the tidal debris surrounding the MW at $z > 50$~kpc. Fig.~\ref{fig:sim_poles_1direction_d100}, which only considers material at $d < 100$~kpc, covers the maximum Galactocentric distance available in the observed sample. From this, it is possible to qualitatively infer that a preferential clustering of the orbital poles around the VPOS should not be expected at this distance. Only when material at higher Galactocentric distances is considered ($d \gtrsim 150$~kpc) can the pole overdensity be more clearly appreciated.

\subsection{Statistical analysis of the orbital pole distribution with mock catalogues}
\label{stat_analysis}
Assuming that future surveys will be able to identify and provide information on more distant streams, we can statistically estimate the Galactocentric distance at which the anisotropic distribution of orbital poles around the VPOS should be significantly preferred over a homogeneous one in the fly-by scenario. For this purpose, we applied the statistical analysis described in Section~\ref{Obs_analysis} to mock catalogues constructed with the simulation data of \citet{Banik_2022}. To create these catalogues, we subtracted all the particles and gas cells at $z < 50$~kpc in order to remove the simulation elements that did not correspond to the tidal debris of the interaction. From the resulting dataset, we randomly selected $N$ elements (gas and particles) and obtained their orbital pole direction. We then obtained their Bernoulli probability ($P$) for a homogeneous distribution within a conical area centred on ($l_{\textrm{VPOS}}$, $b_{\textrm{VPOS}}$) = ($164.0^{\circ}$, $-6.9^{\circ}$) and aperture angle of $36.87^{\circ}$. We repeated this procedure 1000 times for each case and we selected the mode of $P$ as the nominal value, and the $1\sigma$ standard deviation (obtained by fitting the $P$ distribution to a Gaussian) as the error.

\begin{table*}
	\caption{Bernoulli probability for the simulation sample.}
	\centering
	\resizebox{\textwidth}{!}{%
	\begin{tabular}{c|c|c|c}
		 Distance cut  & $P$ ($N=100$)     & $P$ ($N=300$)   &  $P$ ($N=500$)  \\ \hline
		$d<100$~kpc & $0.92^{+0.00}_{-0.19}$ ($(0.09^{+0.24}_{-0.00})\sigma$) & $0.98^{+0.00}_{-0.19}$ ($(0.03^{+0.24}_{-0.00})\sigma$)  & $0.98^{+0.00}_{-0.01}$ ($(0.02^{+0.02}_{-0.00})\sigma$) \\
		$d<150$~kpc & $(2.41^{+14.80}_{-0.00})\times 10^{-2}$ ($(2.26^{+0.00}_{-0.89})\sigma$) & $(3.74^{+13.39}_{-0.00})\times 10^{-3}$ ($(2.90^{+0.00}_{-0.52})\sigma$)  & $(5.08^{+15.01}_{-0.00})\times 10^{-4}$ ($(3.48^{+0.00}_{-0.39})\sigma$) \\
		$d<200$~kpc & $(1.78^{+10.72}_{-0.00})\times 10^{-2}$ ($(2.37^{+0.00}_{-0.83})\sigma$) & $(1.99^{+7.81}_{-0.00})\times 10^{-3}$ ($(3.09^{+0.00}_{-0.51})\sigma$)  & $(2.70^{+8.60}_{-0.00})\times 10^{-4}$ ($(3.64^{+0.00}_{-0.39})\sigma$) \\
		$d<350$~kpc & $(8.43^{+31.07}_{-0.00})\times 10^{-3}$ ($(2.63^{+0.00}_{-0.58})\sigma$) & $(1.36^{+7.34}_{-0.00})\times 10^{-4}$ ($(3.82^{+0.00}_{-0.49})\sigma$)  & $(1.74^{+3.45}_{-0.00})\times 10^{-6}$ ($(4.78^{+0.00}_{-0.22})\sigma$) \\
		 \hline
	\end{tabular}
	}
	\tablefoot{The $P$ values represent the probability that, for a randomly selected sample, a homogeneous pole distribution provides the same (or higher) fraction of VPOS members as the pole distribution of the \citet{Banik_2022} simulation. For example, according to this method, there is a $(0.84^{+3.11}_{-0.00})\%$ chance that, if 100 elements are randomly drawn from a homogeneous distribution with $d<350$~kpc, these have the same or a higher fraction of VPOS members than 100 elements drawn from the anisotropic distribution of \citet{Banik_2022} at this distance. The corresponding number of standard deviations is also indicated in parentheses. This probability was obtained at different upper distance cuts ($d<$) and for different sample sizes ($N$), considering only one possible orbital sense (half $l$-parameter space). In these simulations, a lower distance cut in the z direction of $z>50$~kpc was imposed to remove the disc contamination.}
	\label{tab_P_1direction}
\end{table*}

\begin{table*}
	\caption{Similar to Table~\ref{tab_P_1direction}, but for the full $l$-parameter space.}
	\centering
	\resizebox{\textwidth}{!}{%
	\begin{tabular}{c|c|c|c}
		   Distance cut  & $P$ ($N=100$)     & $P$ ($N=300$)   &  $P$ ($N=500$)  \\ \hline
		$d<100$~kpc & $(4.26^{+6.83}_{-0.00})\times 10^{-2}$ ($(2.03^{+0.00}_{-0.43})\sigma$) & $(2.83^{+26.58}_{-0.00})\times 10^{-2}$ ($(2.19^{+0.00}_{-1.14})\sigma$)  & $(1.86^{+12.76}_{-0.00})\times 10^{-2}$ ($(2.35^{+0.00}_{-0.90})\sigma$) \\
		$d<150$~kpc & $(4.54^{+7.74}_{-0.00})\times 10^{-4}$ ($(3.51^{+0.00}_{-0.27})\sigma$) & $(3.77^{+3.65}_{-0.00})\times 10^{-8}$ ($(5.50^{+0.00}_{-0.12})\sigma$)  & $(3.38^{+6.79}_{-0.00})\times 10^{-14}$ ($(7.58^{+0.00}_{-0.14})\sigma$) \\
		$d<200$~kpc & $(1.96^{+3.87}_{-0.00})\times 10^{-5}$ ($(4.27^{+0.00}_{-0.25})\sigma$) & $(1.25^{+2.99}_{-0.00})\times 10^{-10}$ ($(6.43^{+0.00}_{-0.19})\sigma$)  & $ \lesssim 1.68 \times 10^{-12}$ $((\gtrsim 7.06 \sigma))$ \\
		$d<350$~kpc & $(1.14^{+2.60}_{-0.00})\times 10^{-6}$ ($(4.87^{+0.00}_{-0.24})\sigma$) & $(7.36^{+15.62}_{-0.00})\times 10^{-16}$ ($(8.06^{+0.00}_{-0.14})\sigma$)  & $\lesssim 1.80 \times 10^{-20}$ $((\gtrsim 8.54 \sigma))$ \\
		 \hline
	\end{tabular}
	}
	\tablefoot{The last two lines in column $P(N = 500)$ are only given as approximate values due to difficulties of the code in handling extremely low probabilities in a long-tailed probability distribution.}
	\label{tab_P_2directions}
\end{table*}

In Table~\ref{tab_P_1direction} we show the resulting $P$ values and their errors. In this case, we consider only one orbital sense direction; therefore, our selected VPOS area covers $20\%$ of the parameter space. These $P$ values were obtained for different $N$ values and at different upper distance cuts.

If future surveys also provide enough information on the stellar streams to discern the orbital sense of the poles, it would be possible to obtain the $P$ value for the full parameter space in the $l$ direction. Given that, in this case, the selected VPOS area would enclose $10\%$ of the parameter space, finding a large number of poles clustered around the VPOS would yield a higher statistical significance in favour of the anisotropic distribution with respect to the previous case. We repeated the analysis for the full parameter space and we show the resulting $P$ values and their errors in Table~\ref{tab_P_2directions}.

Both Table~\ref{tab_P_1direction} and Table~\ref{tab_P_2directions} show a steep decrease in the $P$ values between the `$d < 100$~kpc' and the `$d < 150$~kpc' distance cuts. This suggests that $d \approx 150$~kpc might be the distance at which a significant clustering of the orbital poles around the VPOS starts to become appreciable.

\section{Discussion}
\label{discussion}

Previous studies on the stellar streams of the MW have discussed different effects that could be disturbing the orbits of the streams; for example, stream interactions with other substructures orbiting the MW \citep{Brooks_2025} and with elements of the MW disc \citep{Thomas_2023}, precession and nutation of their orbital poles due to the asphericity of the MW potential \citep{Erkal_2016}, or, in a $\Lambda$CDM context, disturbances from dark matter substructures around the MW \citep{Delos_2022}. In the MONDian simulations of \citet{Banik_2022}, we found that, shortly after the interaction, there is a clear orbital pole clustering very close to the current VPOS position at d < 100 kpc (see Appendix~\ref{App_t06}). In the present day, the simulated poles do not display such a clear overdensity close to the VPOS at low Galactocentric distance. But the pole clustering around this region becomes (progressively) significant again with increasing distance. From this, we infer that the effect that caused the dispersion of the original pole overdensity is likely to be dependent on the distance. Of all the effects mentioned above, precession is the one with the most clear dependency on distance in its modelling (see Eq.~\ref{eq_precess}). We therefore investigate its effect on the orbital poles in this section.

Since the MW potential is not perfectly spherical, it is expected that the substructures orbiting it will experience a torque that causes their orbits to precess. This means that the orbital poles of these substructures will progressively tilt away from their original pole positions in the (Galactocentric) $l$ direction $-$ that is, the horizontal direction in Figures~\ref{poles_galstreams} and \ref{fig:sim_poles_1direction}.

\citet{Ibata_2001} estimated that the precession rate for a particle orbiting a flattened logarithmic potential is given by\footnote{We note that, although Eq.~\ref{eq_precess} was derived for a Newtonian potential with a dark matter halo, \citet{Read_2005} and \citet{Thomas_2017} found that the orbits of the Sagittarius dwarf galaxy and its stellar stream in a MONDian potential could barely be distinguished from their corresponding orbits in a $q \approx 0.9$ dark matter halo. Therefore, Eq.~\ref{eq_precess} should still be approximately valid for substructures within the MOND regime.}
\begin{eqnarray}
	\omega_p(r) = -\frac{3}{2}\left(\frac{v_o}{d}\right)\frac{(1-q^2)}{(1+2q^2)}\cos (\psi_0) \, ,
\label{eq_precess}
\end{eqnarray}
where $v_0$ is the circular velocity of the particle orbiting the potential, $d$ is its distance to the potential centre, $\psi_0$ is the angle of the orbital pole with the $z$ axis ($\psi_0 = 90^{\circ} - b$), and $q$ is the axis ratio of the potential ($q<1$ for oblate potential shapes and $q>1$ for prolate potential shapes). Since this equation was derived under the assumption of $q\approx 1$, its accuracy decreases as $q$ departs from the quasi-spherical case; for example, at $q=0.8$ it is expected to be accurate to the first decimal digit \citep{Ibata_2001}.

Taking $v_0 = 220$~km/s \citep{Sofue_2020} and $q=0.9$ \citep{Huang_2024}, we find that substructures near the VPOS normal ($b \approx -7^{\circ}$) that have been orbiting the MW potential for $\approx 7$~Gyr should have precessed by $\approx 60^{\circ}$ at 20~kpc, $\approx 24^{\circ}$ at 50~kpc, and $\approx 12^{\circ}$ at 100~kpc. For substructures that are at a slightly higher $|b|$ (e.g. $|b| = 15^{\circ}$), the expected precession is higher ($\approx 127^{\circ}$ at 20~kpc, $\approx 51^{\circ}$ at 50~kpc, and $\approx 25^{\circ}$ at 100~kpc). Given the strong differences in precession rate values for substructures with different $b$ and $d$ values, this effect should have had a major influence on the dispersion of the pole overdensity. 

By using Eq.~\ref{eq_precess}, we are able back-trace the orbital poles of the present-day snapshot of the simulation to their positions $6.6$~Gyr ago (the time at which we observe a strong pole overdensity in the simulation) in terms of precession effects. With this, we are able to asses the effect of precession on the simulated sample. Fig.~\ref{fig:sim_poles_1direction_precess} shows the pole distribution of the back-traced sample for different distance cuts, similarly to Fig.~\ref{fig:sim_poles_1direction}. Table~\ref{tab_P_backtraced} shows the results of performing the statistical analysis described in Section~\ref{stat_analysis} on the back-traced sample. From these, it can be appreciated that, once precession effects are reversed (approximately, as Eq.~2 is just an analytical equation), the orbital clustering within the VPOS region becomes more statistically significant for all distance cuts. In particular, we note that the back-traced distance cut $d < 100$~kpc sample could plausibly display a VPOS-region overdensity with $2\sigma$ confidence already. We also note that this distance cut is likely to have a larger fraction of elements that were close or connected to the MW disc during the formation and evolution of the tidal tails (see Appendix~\ref{App_t06}). This, in addition to precession, could also contribute to there being a lower fraction of VPOS members at low Galactocentric distances.

Regarding the real (observed) stream sample, we have less information with which to determine the reasons for its current orbital pole distribution. For example, we barely have any streams at $d > 50$~kpc, we do not know how the distribution looked like a few gigayears ago, and we do not know how long the streams and their progenitors have been influenced by the MW potential for.

Still, we can apply Eq.~\ref{eq_precess} to the pole positions of the streams in the observed sample in order to estimate their positions 6.6~Gyr ago. Repeating the analysis described in Section~\ref{Obs_analysis} on the sample of back-traced orbital poles, we find that the fraction of VPOS members is now 0.25 for the full sample (91 streams), which reduces the $P$ value to $0.13 \pm 0.04$ (in the original present-day sample $P = 0.83 \pm 0.09$).

An additional complication of the observed sample is the uncertainty in the orbital pole position of a stream. Streams that present very large uncertainties (that is, very extended pole tracks) are likely to have had their internal structure perturbed throughout their trajectories. Even if, as we argue in this study, precession is one of the main mechanisms responsible for displacing the orbital poles of the streams from their original angular positions, this does not necessarily mean that it is the only or the main mechanism responsible for perturbing the streams' internal structure (at least not for all streams). For example, a stream at a low Galactocentric distance with $b = 60^{\circ}$ will be displaced from its original pole position by quite a lot but, if most stars composing this stream have similar $b$ values, the internal structure of the stream will not be so affected by precession, as the stars in the stream will be precessing at similar rates. We also note that precession only causes displacements in the $l$ direction, so the uncertainty in the $b$ direction observed for some of the streams in the catalogue must have been caused by a different phenomenon (e.g. nutation, interaction with other substructures, or interaction with disc components).

If a stream's internal substructure were perturbed by a different phenomenon in the last few gigayears and this caused it to have its pole track extended in the $b$ direction, back-tracing all poles in its pole track by 6.6~Gyr using Eq.~\ref{eq_precess} $-$ as we did in the previous experiment $-$ will lead us to obtain unrealistically extended pole tracks in the l direction. As this implies a very large uncertainty in the pole position of the stream, we repeated the experiment, removing the back-traced streams with very large uncertainties (occupying more than half of the $l$ parameter space). This left us with 50 streams. This sample has a fraction of 0.32 VPOS members, which reduces the $P$ value even further ($P = 0.03 \pm 0.03$).

Besides this, the observed pole distribution also exhibits other properties consistent with the scenario of an initially anisotropic distribution disturbed by precession. First, the pole gap around the $b = 0^{\circ}$ direction (see Fig.~\ref{poles_galstreams} and Fig.~\ref{poles_square}) could be an indication that a significant fraction of the streams has been strongly affected by precession. This is because precession can also widen the streams \citep{Ibata_2001, Helmi_2004, Johnston_2005, Erkal_2016}. This phenomenon takes place due to the fact that the stars composing a stream can have different angular positions (in $b$), and thus slightly different precession rates. As the orbital plane of some of the stars precesses somewhat faster or slower than the orbital plane of other stars in the stream, the angular spread of the stream increases with time. This can make the stream harder to observe $-$ as it becomes more diffuse $-$ or get destroyed if the widening is very significant. The widening effect becomes stronger towards polar orbits \citep[see][eq.~26]{Erkal_2016}, which would explain the lack of poles around this direction. Interestingly, most of the few streams that can be observed around $b \approx 0^{\circ}$ are concentrated in the regions $120^{\circ} < l < 180^{\circ}$ and $270^{\circ} < l < 300^{\circ}$. This could be related to the proximity of the VPOS and the Sagittarius dwarf galaxy (red star in Fig.~\ref{poles_galstreams}) to these regions, and could imply a direct association of the nearly polar streams with these substructures, as well as a recent formation time for these streams.\footnote{\citet{Pawlowski_2012} also noted that there is a detection limit for streams whose heliocentric distance ($d_{\textrm{helio}}$) is smaller than $d_{\odot-\textrm{GC}}$ around $(l, b) = (180^{\circ}, 0^{\circ})$ (see their section 3.5). This would remove some small $d_{\textrm{helio}}$ streams close to this angular position. However, this is insufficient to explain the scarcity of quasi-polar streams at $180^{\circ} < l < 270^{\circ}$ for all distance ranges.}

Second, the statistical analysis in Section~\ref{Obs_analysis} showed a progressive increase in the pole fraction around the VPOS when the sample was reduced to streams with higher Galactocentric distances. The preference for the VPOS direction could still not be proven at high significance, but, according to the results shown in Section~\ref{stat_analysis}, this should not be possible until material to at least $d \approx 150$~kpc is also considered within the sample.

Even though streams at large Galactocentric distances are currently difficult to observe, larger objects such as satellite galaxies can already be used to test for a preferential pole direction at larger distances. For instance, \citet{Pawlowski_2020} used the 11 classical satellites of the MW \citep{Metz_2007}, whose proper motions are already available, to test for a correlation between their orbital poles. They found that, among these, eight satellites aligned to $<20^{\circ}$ with a common direction, and seven even orbited in the same sense. The co-orbiting satellites are at high Galactocentric distances ($\approx$ 50-240~kpc) and have quasi-polar orbits ($b \lesssim 20^{\circ}$), which would have made their precession rates slower. Because of this, it is plausible that these objects were able to approximately conserve their original pole orientation $-$ unlike most of the streams, which are mainly at $d \leq 20$~kpc. The sample used by \citet{Pawlowski_2015}, which consisted of 71 stellar objects with distances up to 365~kpc (including the satellite galaxies), also displayed a good agreement between the orbital orientations of most of its objects. Very recent studies are also finding new streams at somewhat higher galactocentric distances (d > 25~kpc) and quasi-polar orbits which are also within the VPOS region \citep{Tian_2025}. In this regard, we consider the observations to still be in good agreement with the expectations of the VPOS hypothesis.

\section{Conclusions}
\label{conclusions}
Following the same procedure as the studies of \citet{Pawlowski_2012} and \citet{Riley_2020}, we have analysed the orbital pole distribution of the streams in the recently updated \textsc{galstreams} catalogue. In our statistical analysis, we obtained the fraction of streams that have orbital poles close ($<36.87^{\circ}$) to the VPOS normal. We then calculated the Bernoulli probability of obtaining such a fraction, assuming that the orbital pole distribution of the pole sample is homogeneous. With this, we aimed to distinguish between a scenario in which the streams have been accreted into the MW in an approximately isotropical way (e.g. filamentary accretion, group infall), and a scenario in which most of the streams have a common origin (e.g. the tidal tails resulting from a past MW-M31 interaction) and should therefore present a preferred orbital pole direction. The former is expected in a $\Lambda$CDM scenario, while the latter can only be explained currently with a MOND-like model.

The results of our statistical analysis show that the orbital pole distribution of the streams is highly consistent with a homogeneous distribution when the whole sample is considered. This seems to support the isotropic accretion scenario. However, when we repeated the analysis subtracting the streams at low Galactocentric distance ($d < 5$~kpc, $d < 10$~kcp and $d<15$~kpc), we found a decreasing trend on the probability for a homogeneous distribution (although still at $<2\sigma$ significance). This trend could be explained if the stream poles originally had a more anisotropic distribution but precession effects $-$ which are stronger for lower Galactocentric distances $-$ had displaced the stream poles. Indeed, when we back-traced (using Eq.~\ref{eq_precess}) the pole positions of our observed sample to their past positions 6.6~Gyr ago $-$ that is, to an epoch in which they are not expected to be significantly affected by precession yet $-$ and repeated the analysis, we found that the distribution became even more anisotropic around the VPOS direction, which further supports this hypothesis.

We also noted that precession effects would have caused the streams to fan out with time, which  would end up destroying them or making them harder to observe. The lack of streams near polar orbits $-$ where the fan out effect should be stronger $-$ can be considered an indication that the streams are indeed strongly affected by precession. The few streams observed in these orbits are either close to the VPOS normal or to the Sagittarius dwarf galaxy's orbital pole. We hypothesise that the origin of these streams might correlate to these substructures. Given that many of these streams are fairly close to the Galactic centre ($d < 10$~kpc), we presume that they have not been orbiting the MW for a very long time, since otherwise they would have widened too much to be observed.

By using hydrodynamical simulations of a MW-M31 interaction, we confirmed that no clear signs of pole clustering should be expected for material at $d < 100$~kpc $-$ the Galactocentric distance within which all our sample is contained. This can be attributed largely to the precession of the orbital poles, but also to the fact that several elements of this sample might have been connected to the Galactic disc at an earlier epoch (so that their orbit has also been influenced by the motion of the disc).

With the simulation data, we generated mock catalogues for different distance cuts, and we applied the statistical analysis used for our observed sample. This gave us an estimation of the Galactocentric distance at which the distribution should be significantly distinguishable from a homogeneous one.

From these results, we infer that a sample including material at $d \approx 150$~kpc should already manifest the anisotropy around the VPOS normal with moderate significance ($\approx 2\sigma$). This significance level could be improved by having a larger sample ($N > 100$), by including information on the orbital sense of the stream, or by subtracting the streams at small Galactocentric distances from the sample.

Even though we have not discussed the orbital pole distribution of the globular cluster population in this work, the conclusions drawn from the simulated data should also apply to these objects. In the catalogue used by \citet{Riley_2020}, the globular clusters classified as `old halo' and `bulge/disc' population have $d < 90$~kpc \citep[see figure 1 in][]{Mackey_2005}, and their orbital poles do not present a significant clumping in the direction of the VPOS normal. On the other hand, the young halo population has a larger fraction of clusters at larger Galactocentric distances (with $d \lesssim 120$~kpc), and the \citet{Riley_2020} results show that this population has a significantly higher fraction of VPOS members ($\approx 0.3$) with respect to the others. This is in agreement with the results of \citet{Pawlowski_2012} and with our previous remarks about structures at higher distances presenting a higher degree of clustering around the VPOS. Still, we also note that the old halo globular clusters and the bulge and disc population are proposed to have formed with the early MW, while the young halo globular clusters are thought to have entered the MW potential later \citep{Zinn_1993, Parmentier_2000, Pawlowski_2012}. Having a different origin $-$ besides precession effects $-$ would also explain why these subgroups of globular clusters do not share a common preferential pole direction. This could also be the case for some of the streams. However, based on \citet{Bonaca_2021}, the observable streams whose progenitors are globular clusters formed with the MW should be a minority.

Verifying that the MW streams and the young halo globular clusters are indeed forming, together with the plane of satellites, a VPOS can be of great relevance to understanding the history and formation processes of our Galaxy. Among the possible scenarios that could have led to the formation of such a structure, the MW-M31 fly-by scenario can be regarded as the most promising one, since it explains not only the VPOS formation but also the warp of the MW disc and the relative inclination of the MW and the M31 discs \citep{Bilek_2018}. In addition to this, a recent study by \citet{Kroupa_2024} also suggests that the putative Keplerian decline of the outer rotation curve of the Galaxy \citep{Sylos_Labini_2023, Jiao_2023, Wang_2023, Ou_2024} may be due to a Galactic-disc-perturbing fly-by.

While the current data on streams and young halo globular clusters is still insufficient to confirm that the majority of their population is associated with the VPOS, it already shows promising results in favour of this hypothesis. We expect that future surveys, which can identify more distant streams and globular clusters, will allow us to confirm or reject, in the coming years, the hypothesis of a preferential orbital pole direction for most MW substructures.

\section*{Data availability}
The data used here has been cited and is available in published form.

\begin{acknowledgements}

The authors would like to thank the referee for valuable comments and suggestions that significantly helped to improve this manuscript. EA acknowledges support through a teaching assistantship by the Helmholtz-Institut für Strahlen- und Kernphysik. PK acknowledges support through the DAAD-Eastern Europe exchange programme. IT acknowledges support through a fellowship from the SPODYR group. EA is grateful to Dr. Cecilia Mateu for answering her questions about the \textsc{galstreams} catalogue. EA is also grateful to Dr. Jan Pflamm-Altenburg and Dr. Marcel Pawlowski for feedback and advice regarding this project. The authors would like to thank the members and collaborators of the SPODYR group for useful comments and discussions regarding the topic of this project. The authors acknowledge the use of the \textsc{healpy} and \textsc{HEALPix} package for obtaining the figures and results that require an equal area subdivision of the parameter space (e.g. figures with contour plots).

\end{acknowledgements}

\bibliographystyle{aa}
\bibliography{MW_streams_bbl}

\begin{appendix}

\section{Distance distribution of the streams in the \textsc{galstreams} catalogue}
\label{App_distance}

\begin{figure}[hbt!]
	\centering
	\includegraphics[width = 8.5cm]{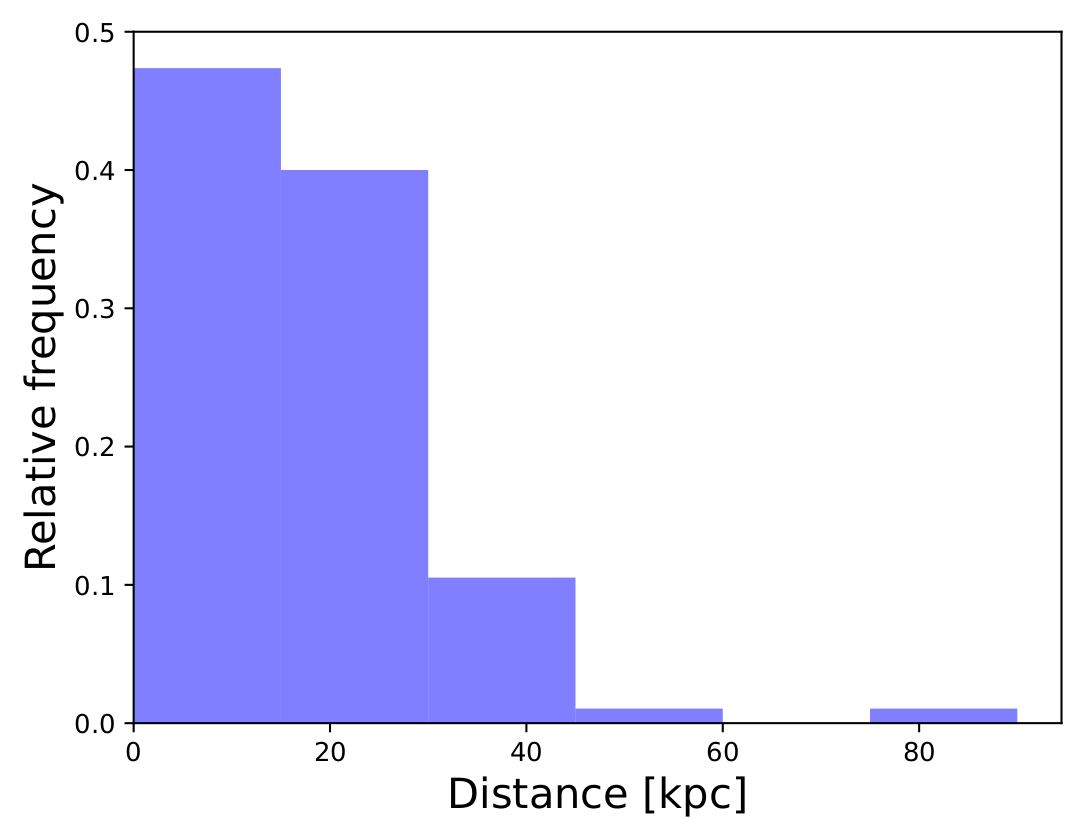}
	\caption{Distance distribution of the stream's nominal poles in bins of $15$~kpc.}
	\label{hist_distance}
\end{figure}

A histogram with the distance distribution of the streams included in the \textsc{galstreams} catalogue is shown in Fig.\ref{hist_distance}.

\FloatBarrier
\section{Distribution of the stream poles}

\begin{figure}[hbt!]
	\centering
	\includegraphics[width = 8.5cm]{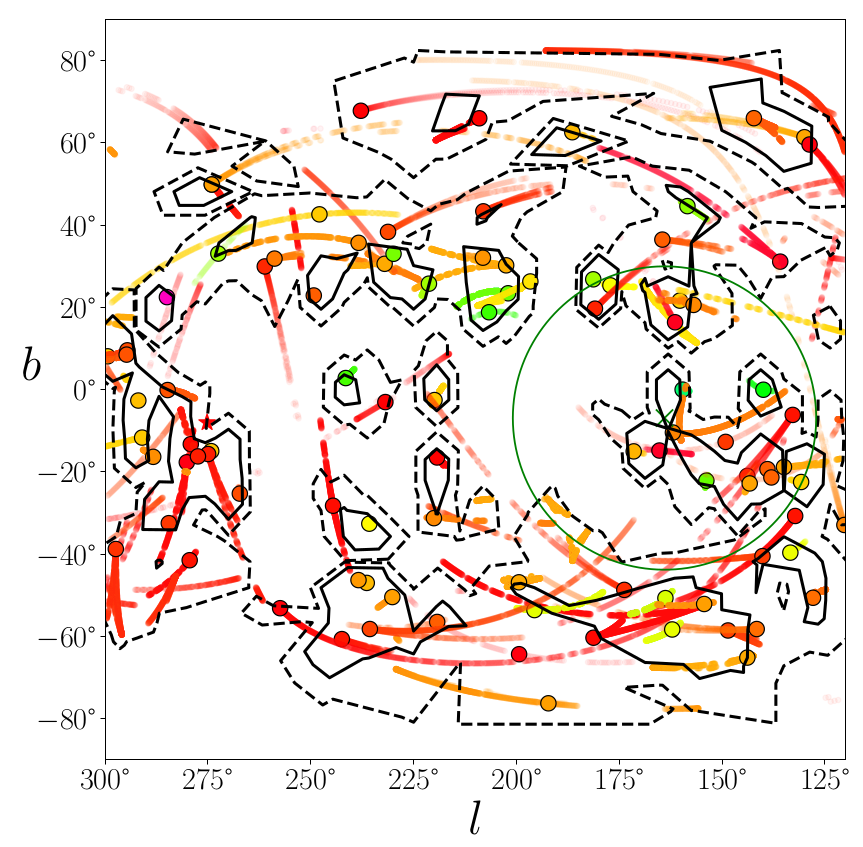}
	\caption{Figure similar to Fig.~\ref{poles_galstreams} but for a 2D projection. This figure also includes contour plots that enclose different stream pole areas as a function of their pole fraction. Solid lines correspond to a $1\sigma$ contour level and dashed lines correspond to a $2\sigma$ level in this plot. For obtaining the fraction of poles per bin, we divided the parameter space in 384 bins of equal area and we distributed the statistical `weight' of each stream pole equally among its uncertainty pole track.}
	\label{poles_square}
\end{figure}

Fig.~\ref{poles_square} shows a plot equivalent to Fig.~\ref{poles_galstreams}, but for a 2D square projection and contour plots highlighting the regions with higher pole fraction. This verifies that the lack of orbital poles near polar orbits, which can be appreciated in Fig.~\ref{poles_galstreams}, is not an effect of the Mollweide projection and that there is indeed a lower density of stream poles $-$ with respect to the surrounding regions $-$ in the central part of the figure.

\FloatBarrier
\section{Comparison with previous studies}
\label{App_prev_studies}
In order to accurately compare the results of the different studies that analyse the degree of clustering of the stream poles around the VPOS direction, it is important to account for the differences in the data and the assumptions taken in these studies.

\begin{figure}
	\centering
	\includegraphics[width = 8.5cm]{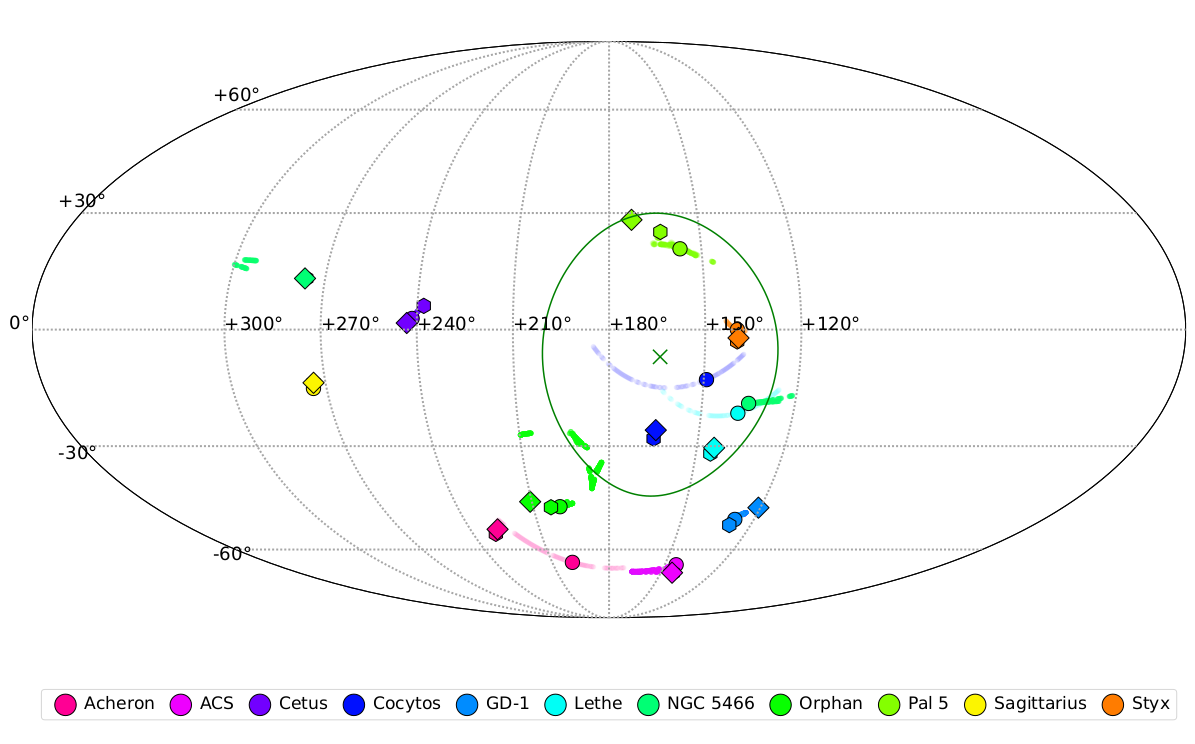}
	\caption{Orbital poles of the streams considered in \citet{Pawlowski_2012} (hexagons) that are also included in the \citet{Riley_2020} (diamonds) and in the \textsc{galstreams} (circles) catalogue. Stream poles corresponding to the same stream in different catalogues are represented with the same color. The $1\sigma$ uncertainty is also included for the \textsc{galstreams} orbital poles.}
	\label{comp_poles}
\end{figure}

In Figure~\ref{comp_poles} we show the orbital poles for the streams that appear in the original study of \citet{Pawlowski_2012}, which also appear in the later study of \citet{Riley_2020} and in the \textsc{galstreams} catalogue considered in our study. The positions of most orbital poles have varied very little with the updated catalogues. The stream whose orbital pole position has changed the most, NGC 5466, has actually moved closer to the VPOS area, instead of further away from it. Therefore, the differences between the results of \citet{Pawlowski_2012} and those of \citet{Riley_2020} and ours are not caused by measurement errors in the orbital pole position of earlier catalogues. However, we also note that some of the streams considered in \citet{Pawlowski_2012} do not appear in the later catalogues due to various reasons: e.g. GCN, a VPOS member in \citet{Pawlowski_2012}, was excluded in the posterior catalogues due to its tentative nature (besides being a gaseous stream); the Magellanic stream, also a VPOS member, was not included in the \textsc{galstreams} catalogue due to being a gaseous stream instead of a stellar stream; NGC 5053, a VPOS stellar debris stream, was not included in \textsc{galstreams} as it was not exactly a stellar stream and was also not very clear.

Another difference between our analysis and those of \citet{Pawlowski_2012} and \citet{Riley_2020} is the assumed position of the VPOS normal. \citet{Pawlowski_2012} used the VPOS normal co-ordinates obtained in \citet{Kroupa_2010} ($l$, $b$) = ($156.4^{\circ}$, $-2.2^{\circ}$) by fitting 24 satellite galaxies, \citet{Riley_2020} used the updated VPOS normal ($l$, $b$) = ($169.3^{\circ}$, $-2.8^{\circ}$) from \citet{Pawlowski_2013}, and, in this work, we use a more recent VPOS estimation, fitted with 38 satellite galaxies ($l$, $b$) = ($164.0^{\circ}$, $-6.9^{\circ}$) \citep{Pawlowski_2015}. As the newly discovered MW satellites are also aligned with the plane of satellites, the new fit to the VPOS normal did not change much with respect to their previous estimations. Because of this, we do not expect that using a slightly different VPOS normal will significantly affect our results. In Table~\ref{tab_obs_statistics_VPOS} we show this by repeating the analysis described in Section~\ref{Obs_analysis} and obtaining $P$ for the VPOS normals of \citet{Pawlowski_2012}, \citet{Riley_2020}, and the one assumed in this study.

\begin{table}
	\caption{Bernoulli probability for the observed sample assuming three different VPOS normal co-ordinates.}
	\centering
	\begin{tabular}{c|c|c|c}
		Distance cut & $P$ (Pawl.)     & $P$ (Ril.)           & $P$        \\ \hline
		\multicolumn{1}{c|}{$d>0$~kpc}     & $0.76 \pm 0.09$ & $0.76 \pm 0.09$ & $0.83 \pm 0.09$  \\ \hline
		\multicolumn{1}{c|}{$d>5$~kpc}     & $0.81 \pm 0.09$ & $0.81 \pm 0.09$ & $0.88 \pm 0.10$  \\ \hline
		\multicolumn{1}{c|}{$d>10$~kpc}    & $0.50 \pm 0.08$ & $0.50 \pm 0.08$ & $0.62 \pm 0.09$  \\ \hline
		\multicolumn{1}{c|}{$d>15$~kpc}    & $0.47 \pm 0.10$ & $0.33 \pm 0.08$ & $0.47 \pm 0.10$  \\ \hline
	\end{tabular}
	\tablefoot{This probability is shown for the VPOS normal assumed in \citet{Pawlowski_2012} ($P$ (Pawl.)), in \citet{Riley_2020} ($P$ (Ril.)), and in this study ($P$).}
	\label{tab_obs_statistics_VPOS}
\end{table}

Even though there are some variations on the exact $P$ value depending on the assumed VPOS normal, all the positions considered show that the stream pole distributions with low distance cuts ($d > 0$~kpc and $d > 5$~kpc) are highly consistent with a homogeneous distribution, and that their $P$ value decreases for higher distance cuts ($d > 10$~kpc and $d > 15$~kpc). In this regard, the conclusions drawn from this analysis are the same for all three VPOS normal positions.

\FloatBarrier
\section{Pole distribution 0.6~Gyr after MW-M31 encounter}
\label{App_t06}
\begin{figure}
	\centering
	\includegraphics[width = 8.5cm]{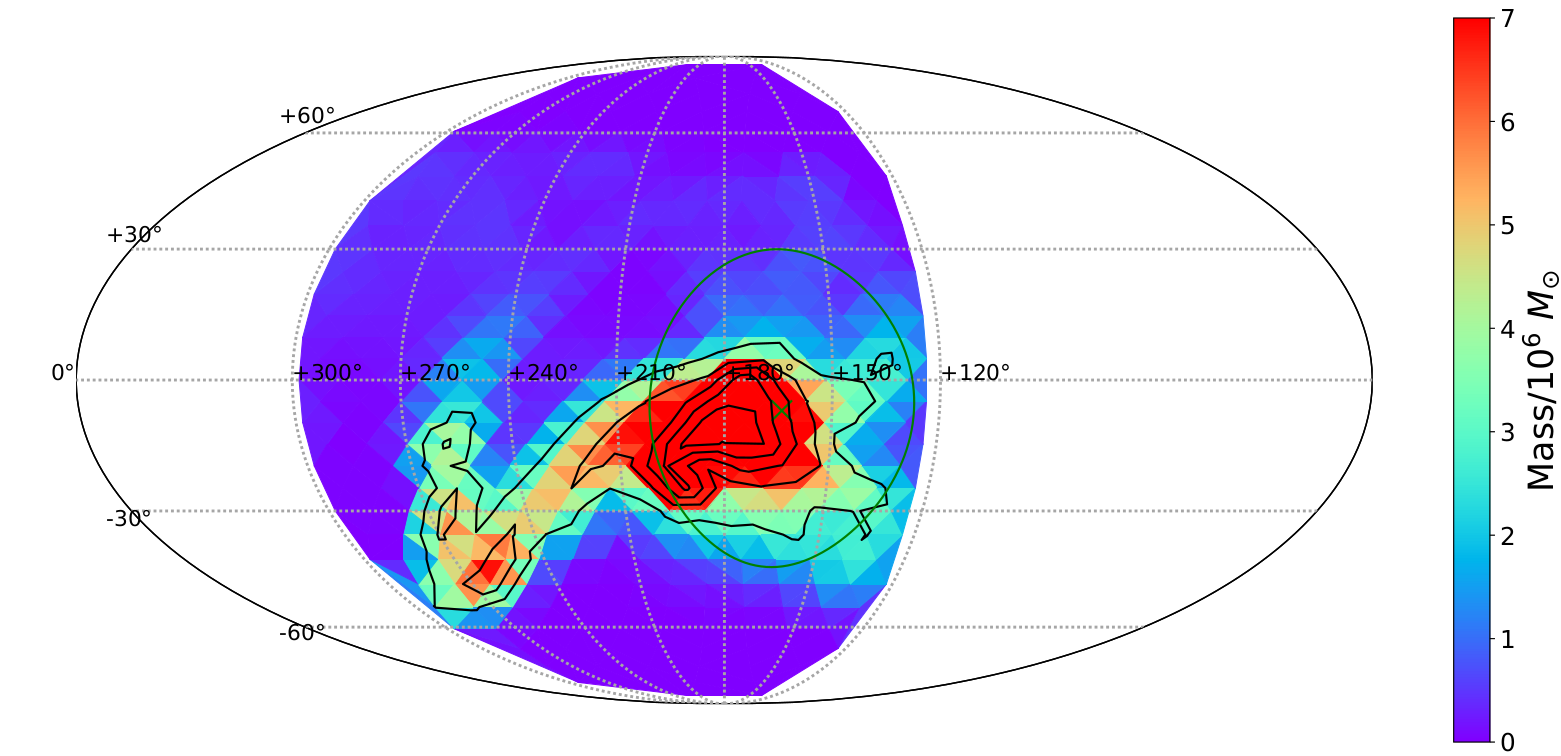}
	\caption{Figure similar to Fig.~\ref{fig:sim_poles_1direction_d100} but for a snapshot taken 0.6~Gyr after the interaction.}
	\label{Sim_50_100_t06}
\end{figure}

In this section we show a snapshot of the orbital pole distribution for a distance cut $d < 100$~kpc at $t=0.6$~Gyr after the MW-M31 interaction in the \citet{Banik_2022} simulation (Fig.~\ref{Sim_50_100_t06}). Here it can be appreciated that, for this epoch, there is a very clear orbital preference towards the VPOS direction even at relatively low Galactocentric distances.

Unlike in Fig.~\ref{fig:sim_poles_1direction}, we do not show the $t=0.6$~Gyr snapshots at higher Galactocentric distances due to the fact that, at this epoch, M31 and its substructures were still very close to the MW (central $d_{\textrm{relative}} \approx 240$~kpc), which already contaminate the MW sample of substructures at $d \gtrsim 150$~kpc.

\begin{figure*}
	\centering
 	\begin{subfigure}[b]{0.475\textwidth}
       \centering
       \includegraphics[width=\textwidth]{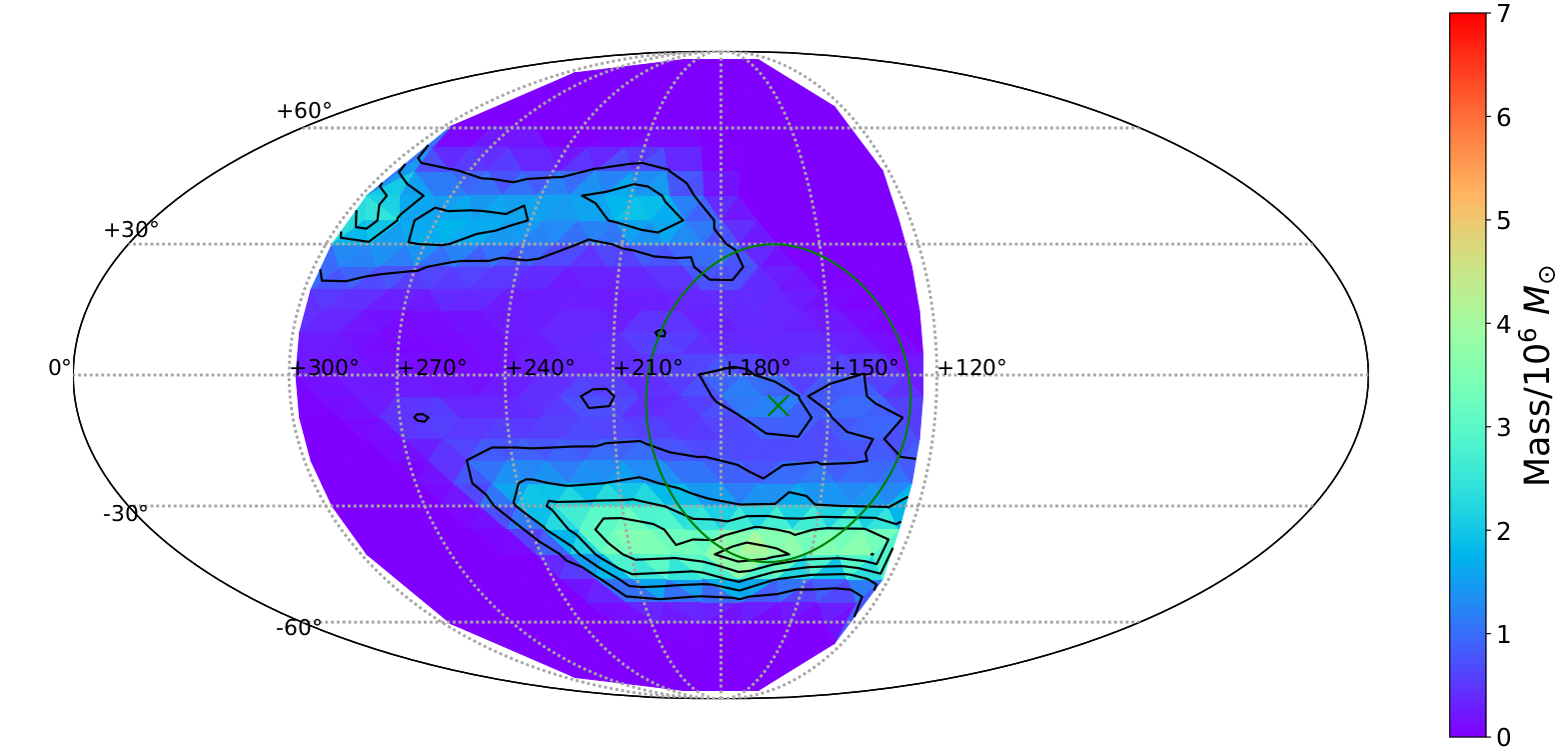}
       \caption[]{{\small Simulated tidal debris at $d<100$~kpc}}    
        \label{fig:sim_poles_1direction_d100_precess}
     \end{subfigure}
     \hfill
     \begin{subfigure}[b]{0.475\textwidth}  
     	\centering 
        \includegraphics[width=\textwidth]{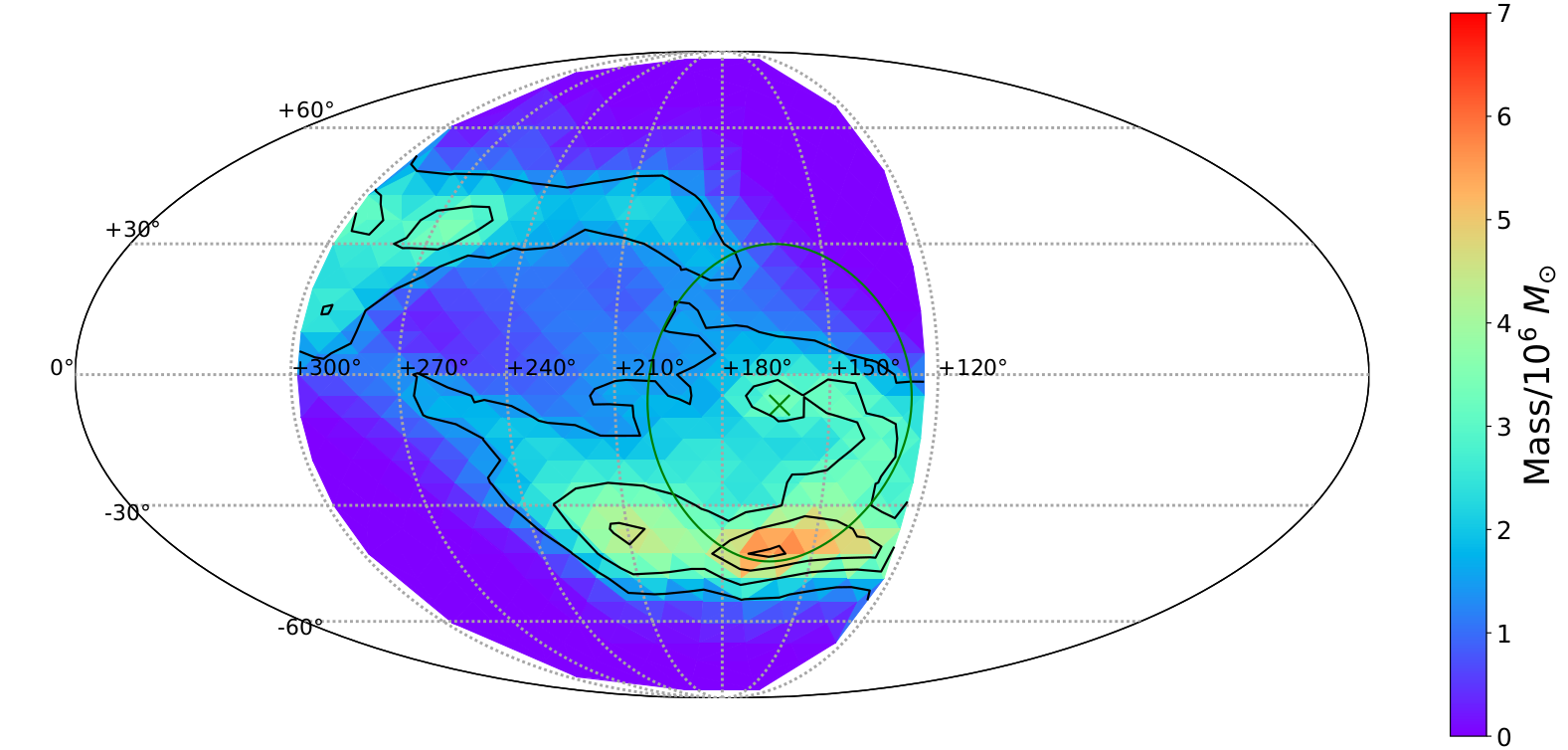}
        \caption[]{{\small Simulated tidal debris poles at $d<150$~kpc}}    
        \label{fig:sim_poles_1direction_d150_precess}
    \end{subfigure}
     \vskip\baselineskip
     \begin{subfigure}[b]{0.475\textwidth}   
     	\centering 
        \includegraphics[width=\textwidth]{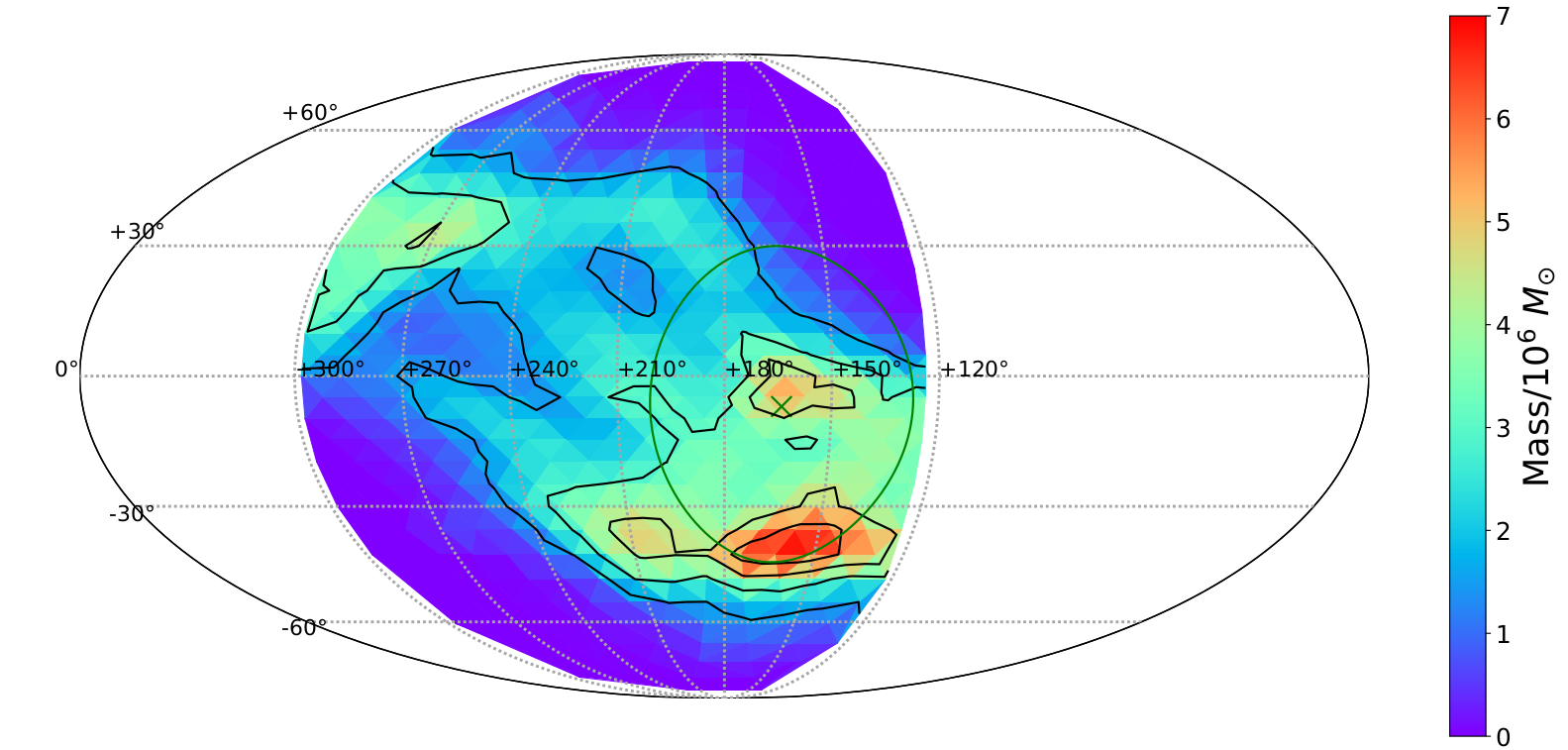}
        \caption[]{{\small Simulated tidal debris poles at $d<200$~kpc}}    
        \label{fig:sim_poles_1direction_d200_precess}
        \end{subfigure}
        \hfill
	\begin{subfigure}[b]{0.475\textwidth}
        \centering 
        \includegraphics[width=\textwidth]{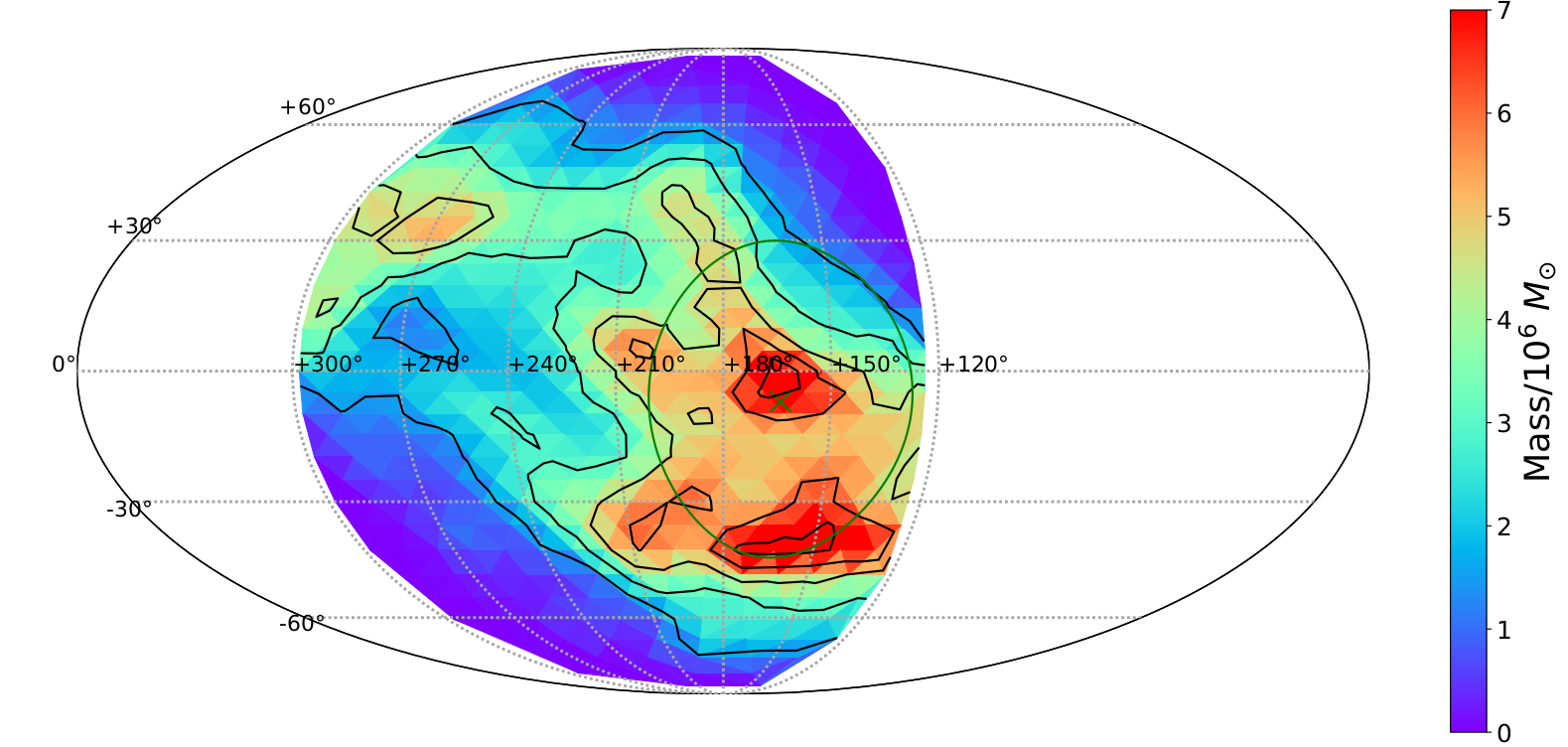}
        \caption[]{{\small Simulated tidal debris poles at $d<350$~kpc}}
        \label{fig:sim_poles_1direction_d350_precess}
    \end{subfigure}
    \caption[]{Figure showing the orbital pole distribution of the particles and gas cells in Fig.~\ref{fig:sim_poles_1direction} after being back-traced for 6.6~Gyr using the precession rate formula (Eq.\ref{eq_precess}). Note that the fraction of elements within the VPOS region increases significantly after being corrected for the effect of precession (see Fig.~\ref{fig:sim_poles_1direction} for comparison).}
    \label{fig:sim_poles_1direction_precess}
\end{figure*}

To evaluate the effect of precession on the simulation elements over the 6.6 Gyr that took place between the $t=0.6$~Gyr snapshot and the present-day snapshot, we back-traced the present-day orbital pole positions applying Eq.~\ref{eq_precess}. Our results (see Fig.~\ref{fig:sim_poles_1direction_precess}) show that, by `undoing' the effects of precession, the clustering of the orbital poles within the VPOS region becomes much more significant. This can also be appreciated in the statistics shown in Table~\ref{tab_P_backtraced}.

Fig.~\ref{fig:sim_poles_1direction_precess} also shows that, even when precession effects are accounted for, the strong pole overdensity shown in Fig.~\ref{Sim_50_100_t06}, at $t = 0.6$~Gyr and $d < 100$~kpc, is no longer visible in the present-day $d < 100$~kpc distance cut (Fig.~\ref{fig:sim_poles_1direction_d100_precess}). One of the main reasons for this is that, in the present-day snapshot, the tidal tails have already separated from the Galactic disc, so the tidal debris shown in Fig.~\ref{Sim_50_100_t06} does not necessarily correspond to the present-day tidal debris at similar Galactocentric distances $-$ since a lot of the tidal debris moved to higher Galactocentric distances after the separation of the tails from the Galactic disc.

For the $d < 100$~kpc distance cut, we note that most of the orbital poles are distributed in extended bands at $|b| \approx 40^{\circ}$ (see Fig.~\ref{fig:sim_poles_1direction_d100} and Fig.~\ref{fig:sim_poles_1direction_d100_precess}). Even if undoing the precession effect returned most of the overdense region in the bands back into the VPOS region, most of the elements in this distance cut still have $b$ values which are relatively far from the VPOS $b$ component normal ($-6.9^{\circ}$). We hypothesise that the material forming these bands (or, at least, a significant part of this material) was part of the tidal tail connecting the disc with the rest of the tidal tail in the $t=0.6$~Gyr snapshot (at $d < 50$~kpc). As it was influenced both by the motion of the disc and of the tidal tail, its $b$ component is lower than that of the VPOS normal, but the $l$ component of the major overdensity in the pole clump is still approximately aligned with the VPOS $l$ component normal ($164.0^{\circ}$) for the back-traced sample.

Besides the fact that the present-day sample is likely to contain elements which were connected both to the Galactic disc and the tidal tails at an earlier epoch, we note that other mechanisms (e.g. the influence of M31 when it was still close to the MW or the varying gravitational potential of the MW perturbed by the encounter) could have also contributed to make the orbital pole concentration in Fig.~\ref{fig:sim_poles_1direction_d350_precess} look less significant than in Fig.~\ref{Sim_50_100_t06}.

\begin{table*}
	\centering
	\resizebox{\textwidth}{!}{%
	\begin{tabular}{c|c|c|c}
		   Distance cut  & $P$ ($N=100$)     & $P$ ($N=300$)   &  $P$ ($N=500$)  \\ \hline
		$d<100$~kpc & $(4.60^{+7.28}_{-0.00})\times 10^{-2}$ ($(2.00^{+0.00}_{-0.45})\sigma$) & $(2.35^{+31.44}_{-0.00})\times 10^{-2}$ ($(2.26^{+0.00}_{-1.31})\sigma$)  & $(1.37^{+5.42}_{-0.00})\times 10^{-2}$ ($(2.46^{+0.00}_{-0.64})\sigma$) \\
		$d<150$~kpc & $(1.97^{+8.00}_{-0.00})\times 10^{-3}$ ($(3.10^{+0.00}_{-0.52})\sigma$) & $(1.40^{+2.79}_{-0.00})\times 10^{-6}$ ($(4.83^{+0.00}_{-0.22})\sigma$)  & $(1.27^{+2.29}_{-0.00})\times 10^{-9}$ ($(6.07^{+0.00}_{-0.17})\sigma$) \\
		$d<200$~kpc & $(1.97^{+8.00}_{-0.00})\times 10^{-3}$ ($(3.10^{+0.00}_{-0.52})\sigma$) & $(1.40^{+3.91}_{-0.00})\times 10^{-6}$ ($(4.82^{+0.00}_{-0.27})\sigma$)  & $(1.27^{+2.50}_{-0.00})\times 10^{-9}$ ($(6.07^{+0.00}_{-0.18})\sigma$) \\
		$d<350$~kpc & $(1.12^{+2.18}_{-0.00})\times 10^{-3}$ ($(3.26^{+0.00}_{-0.32})\sigma$) & $(2.32^{+12.13}_{-0.00})\times 10^{-7}$ ($(5.17^{+0.00}_{-0.35})\sigma$)  & $(3.31^{+4.95}_{-0.00})\times 10^{-11}$ ($(6.63^{+0.00}_{-0.14})\sigma$) \\
		 \hline
	\end{tabular}
	}
	\caption{Table similar to Table~\ref{tab_P_1direction} but for the $P$ values corresponding to the back-traced sample shown in Fig.~\ref{fig:sim_poles_1direction_precess}.}
	\label{tab_P_backtraced}
\end{table*}

\end{appendix}

\label{lastpage}
\end{document}